\documentclass{article}

\usepackage{PRIMEarxiv}

\usepackage[utf8]{inputenc} 
\usepackage[T1]{fontenc}    
\usepackage{hyperref}       
\usepackage{url}            
\usepackage{booktabs}       
\usepackage{amsfonts}       
\usepackage{nicefrac}       
\usepackage{microtype}      
\usepackage{lipsum}
\usepackage{fancyhdr}       
\usepackage{graphicx}       
\graphicspath{{media/}}     
\usepackage{amsmath} 
\usepackage{multicol}

\usepackage{bm}
\usepackage{siunitx}
\usepackage{subcaption}
\usepackage{booktabs}

\usepackage{mathrsfs}
\usepackage[table]{xcolor}
\usepackage{array, multirow, boldline}
\usepackage{pst-node}
\usepackage{tikz}
\usepackage{cancel}

\usepackage{multirow}
\usetikzlibrary{calc}
\usetikzlibrary{matrix, arrows.meta}
\usepackage{chngcntr}   
\numberwithin{equation}{section}
\title{Path-Based Approach for Detecting and Assessing Inconsistency in Network Meta-Analysis: A Novel Method}

\author{
  Noosheen R. Tahmasebi \\
  Institute for Medical Biometry and Statistics (IMBI)\\ Faculty of Medicine and Medical Center, University of Freiburg \\
  Freiburg, Germany\\
  \texttt{noosheen.rajabzadehtahmasebi@uniklinik-freiburg.de}
   \And
  Annabel L. Davies \\
  Bristol Medical School \\
  University of Bristol \\
  Bristol, UK
  \And
  Theodoros Papakonstantinou\\
    Institute for Medical Biometry and Statistics (IMBI)\\ 
  Freiburg, Germany\\
  Laboratory of Hygiene, Social and Preventive Medicine and Medical Statistics\\
  School of Medicine, Aristotle University of Thessaloniki\\
Thessaloniki, Greece
\And
Gerta R\"ucker\\
  Institute for Medical Biometry and Statistics (IMBI)\\ Faculty of Medicine and Medical Center, University of Freiburg \\
  Freiburg, Germany
  \And 
  Adriani Nikolakopoulou\\
      Institute for Medical Biometry and Statistics (IMBI)\\ 
  Freiburg, Germany\\
  Laboratory of Hygiene, Social and Preventive Medicine and Medical Statistics\\
  School of Medicine, Aristotle University of Thessaloniki\\
Thessaloniki, Greece
}

\begin{document}
\maketitle

\begin{abstract}
Network Meta-Analysis (NMA) plays a pivotal role in synthesizing evidence from various sources and comparing multiple interventions.  At its core, NMA relies on integrating both direct evidence from head-to-head comparisons and indirect evidence from different paths that link treatments through common comparators.  A key aspect is evaluating consistency between direct and indirect sources. 
Existing methods to detect inconsistency, although widely used, have limitations. For example, they do not  account for differences within indirect sources or cannot estimate inconsistency when direct evidence is absent.

In this paper, we introduce a path-based approach that explores all sources of evidence without separating direct and indirect. We introduce a measure based on the square of differences to quantitatively capture inconsistency, and propose a Netpath plot to visualize inconsistencies between various paths. We provide an implementation of our path-based method within the netmeta R package. Via application to fictional and real-world examples, we show that our method is able to detect and visualize inconsistency between multiple paths of evidence that would otherwise be masked by considering all indirect sources together. The path-based approach therefore provides a more comprehensive evaluation of inconsistency within a network of treatments.
\end{abstract}


\section{Introduction}\label{sec1}

In medical research, randomized controlled trials (RCTs) have been acknowledged as the most reliable method for comparing treatment options.\cite{book:rct} As the collection of available treatments expands, the potential pairwise comparisons grow drastically.\cite{Lumley:incoherence} The assessment of treatment efficacy commonly relies on direct (dir) comparisons in randomized trials. Indirect (indir) comparisons arise when two treatments  have undergone direct comparisons with at least one other common treatment.\cite{Georgia:consistency} A treatment comparison refers to the difference in  outcome effects between two treatments, which can be derived from either direct head-to-head comparison (known as a contrast) or indirectly through common comparators in a network of studies. Network meta-analysis (NMA) offers a more informed estimate by incorporating both direct and indirect evidence. Unlike traditional meta-analysis, which uses only head-to-head trials, NMA combines data from multiple sources, allowing for comparisons between treatments not directly compared in trials, leading to more comprehensive and precise estimates.\cite{Lu&Ades:loopInconsistency}

The inconsistency of a certain treatment comparison is defined as the conflict between different sources of evidence on that comparison.\cite{Georgia:consistency} Typically, this pertains to the conflict between direct and indirect evidence and has been addressed using various terms, including incoherence.\cite{Lumley:incoherence} The presence of inconsistency in the network of evidence  undermines the reliability of conclusions drawn from NMA. Therefore, investigation of consistency is important for making robust, evidence-based medical decisions and policy recommendations.

Inconsistency is currently assessed using methods that evaluate the agreement between direct and indirect evidence by statistically testing for discrepancies. To assess or detect inconsistency in a network, local methods focus on individual parts or subsets of the network, such as specific treatment contrasts or loops within the network.  Meanwhile, global methods evaluate the overall consistency of the entire network, providing a broad view to detect inconsistencies that affect the data or model as a whole. However, both these approaches have limitations. Local methods are often computationally expensive, and can fail to detect more nuanced inconsistencies between different sources of indirect evidence. Moreover, these methods cannot assess the inconsistency on a comparison when there is no direct evidence available. To address these limitations, we propose a method based on the concept of 'paths of evidence'.\cite{Theo:path} Rather than segregating evidence into direct and indirect sources, we treat each indirect estimate separately. In this framework, a direct comparison becomes just one of many possible paths between two treatments. We then employ the concept of inconsistency to judge the coherence among all the paths between two treatments of interest,  providing a thorough assessment of inconsistency beyond just direct versus indirect comparisons. 

In this article, we make two primary contributions: (i) we perform a concise review of established local methods used to detect and assess inconsistency in NMA, and (ii) we propose a novel path-based method to quantify the degree to which different sources of evidence agree when obtaining a specific network estimate. To illustrate and compare the different methods, we use a toy example of a network with four treatments and five observed comparisons. 

In Section \ref{sec2}, we explore existing measures of inconsistency including the local methods loop-specific \cite{Lu&Ades:loopInconsistency} and side-splitting,\cite{Dias:side-splitting} as well as the global design-by-treatment (DBT) interaction  model.\cite{Higgings:design-by-treatment}  We explain when and how these methods fall short, and  compare their performance via application to our toy example. In Section \ref{sec3}, we introduce the path framework and propose an inconsistency measure based on the square of differences. This measure aims to increase the accuracy and interpretability of inconsistency assessment. In Section \ref{sec4}, we apply our method to a real-world example, demonstrating its implementation and the efficacy of the algorithm in effectively handling larger networks. 

\section{Consistency Assumption in NMA}\label{sec2}

Consider a network of $N$ nodes with treatments $\{T_1, \dots, T_N\}$. There are $M = N(N-1)/2$ possible comparisons and $E$ direct comparisons (edges). In the aggregate NMA model,\cite{aggregate_model} the treatment contrast between any pair of treatments $T_i$ and $T_j$ is associated with a summary effect $\theta^{T_iT_j}$ derived from a pairwise meta-analysis of $S^{T_iT_j}$ studies investigating their comparative efficacy. $\boldsymbol{\Theta}$ is an $M \times 1$ vector containing all the summary effects for direct comparisons, with $0$s for the comparisons that are not directly studied.\cite{Gerta:shortest-path} 

\begin{figure*}
    \centering
    \includegraphics[width=0.4\textwidth]{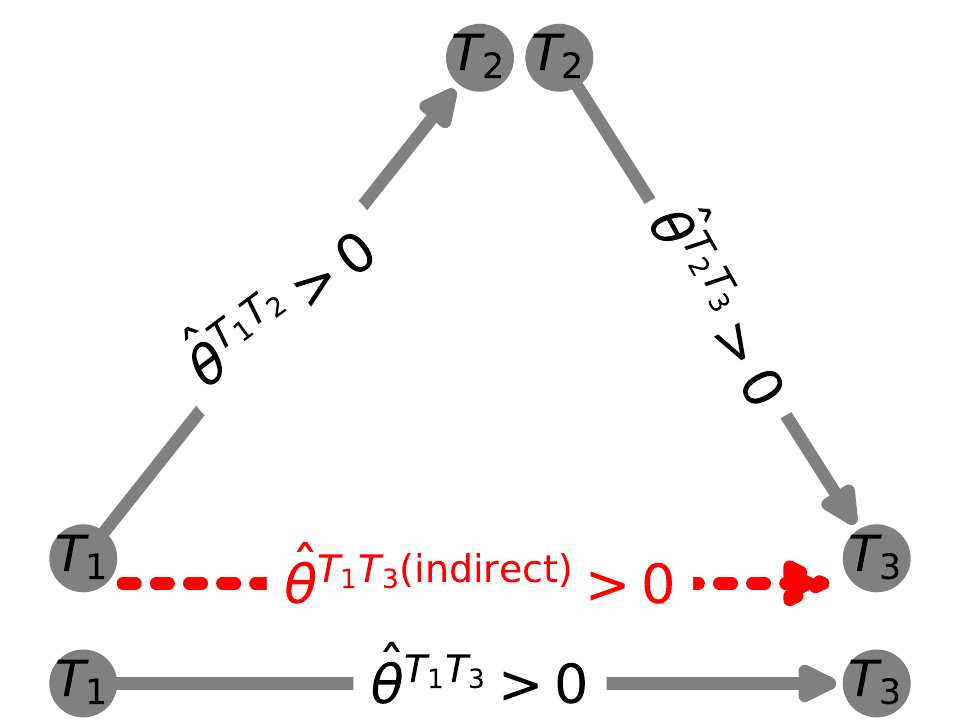}
    \caption{The simplest network with three treatments: $T_1$, $T_2$, and $T_3$. The summary effect for each pair of contrasts, $\hat{\theta}^{T_iT_j}$, is derived from a pairwise meta-analysis based on data from $S^{T_iT_j}$ studies comparing those two treatments. The indirect estimate $\hat{\theta}^{T_1T_3}$ comes from a chain of logic.\label{fig:fig_1}}
\end{figure*}

The reliability of NMA is highly dependent on the consistency assumption, which requires that different sources of evidence for a specific comparison should be in agreement. Let us consider a network with three treatments—$\{T_1, T_2, T_3\}$— where we have the summary effect $\theta^{T_iT_j}$ for every pair  (Figure \ref{fig:fig_1}). We assume a beneficial outcome, such that a positive summary effect ($\theta^{T_iT_j}>0$) indicates that treatment $T_i$ is more effective than treatment $T_j$. If we observe $\hat{\theta}^{T_1T_2} > 0$ ($T_1$ is better than $T_2$) and $\hat{\theta}^{T_2T_3} > 0$ ($T_2$ is better than $T_3$), then the indirect estimate, 
\begin{equation}
    \theta^{T_1T_3\text{(indir)}} = \theta^{T_1T_2} + \theta^{T_2T_3},
    \label{eq:cons}
\end{equation}
implies that $T_1$ should be more effective than $T_3$ ($\hat{\theta}^{T_1T_3} > 0$). 
If the direct summary effect for $T_1T_3$ is negative, $\hat{\theta}^{T_1T_3} < 0$, the consistency assumption is violated. Such an inconsistency between the two  sources of evidence jeopardizes the validity of NMA estimates in this network.

\subsection{Concepts}

We calculate $\hat{\theta}^{T_iT_j\text{(nma)}}$ for the NMA estimate of treatment comparison $T_iT_j$. We collect these estimates in the $M \times 1$ vector $\hat{\boldsymbol{\Theta}}^{\text{nma}}$ calculated via
    \[
    \hat{\boldsymbol{\Theta}}^{nma} = \textbf{\text{H}} \boldsymbol{\Theta}, 
    \] 
where $\textbf{\text{H}}$ is the $M \times M$ full hat matrix.\cite{Gerta:shortest-path} The hat matrix maps the vector of observed summary effects onto the vector of network estimates. Using consistency assumption in Equation \ref{eq:cons}, each element $\hat{\theta}^{\text{$T_iT_j$(nma)}}$ can then be expressed as a linear combination of direct estimates,
\begin{equation}
    \hat{\theta}^{T_iT_j\text{(nma)}} = \textbf{\text{H}}_{T_iT_j,:} \boldsymbol{\Theta}, 
\label{eq:nma_estimate}
\end{equation}
where $\textbf{\text{H}}_{T_iT_j,:}$ is the row of $\textbf{\text{H}}$ corresponding to the comparison $T_iT_j$.

Geometrically, $\hat{\boldsymbol{\Theta}}^{\text{nma}}$ is the projection of $\boldsymbol{\Theta}$ onto an $M$-dimensional subspace.  For any row $\textbf{\text{H}}_{r, :}$, the element $h_{rc}$  indicates the leverage or influence that the column $\textbf{\text{H}}_{:, c}$ exerts on that row. Therefore, the leverages of the direct comparisons in the estimated NMA are indicated by the diagonal elements of the hat matrix, $h_{dd}$.\cite{hat_matrix} For a specific comparison $T_iT_j$, one can obtain a directed  network from node $T_i$ to node $T_j$ using the corresponding row $\textbf{\text{H}}_{r=T_iT_j, :}$. Every element $h_{rc}$ of this row represents an edge $c=T_lT_k$, and the direction of each edge is determined by the sign of $h_{rc}$.\cite{Konig_visualise_flow, Theo:path} That is, if $h_{rc}>0$, the edge points from node $T_l$ to node $T_k$, while $h_{rc}<0$ corresponds to the direction $T_k$ to $T_l$. 

\subsection{Motivating Example}

\begin{figure*}
    \centering
    \includegraphics[width=\textwidth]{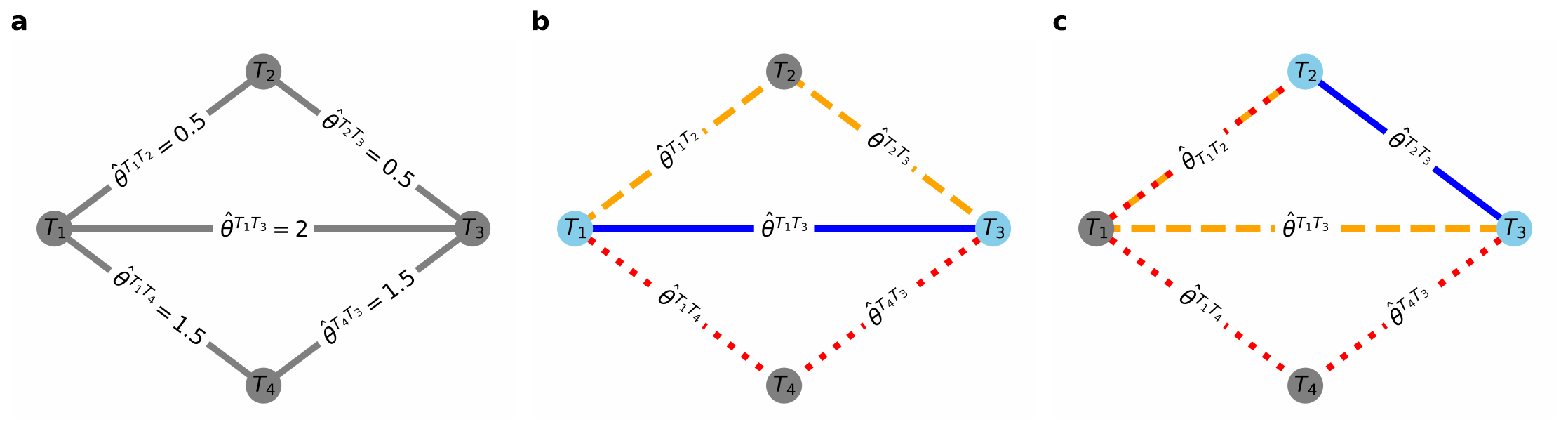}
    \caption{(a) A fictional example with four treatments $\{T_1, T_2, T_3, T_4\}$, and five direct comparisons $T_1T_2$, $T_1T_3$, $T_1T_4$, $T_2T_3$, and $T_3T_4$. The $\hat{\theta}^{T_iT_j}$ refers to the pairwise aggregate effect of the comparison $T_i$ vs $T_j$. Three different paths of evidence for the comparison (b) $T_1$ vs $T_3$ and (c) $T_2$ vs $T_3$. The treatments of interest are shown as blue nodes. The direct evidence is represented by a blue solid line, while the indirect evidence is represented by dotted lines in orange and red. In panel (c), both the red and orange lines pass through the edge connecting $T_1$ and $T_2$.\label{fig:fig_2}}
\end{figure*}

Figure \ref{fig:fig_2}(a) shows an illustrative example network featuring four treatments ($N = 4$, $M = 6$) denoted by $\{T_1, T_2, T_3, T_4\}$, with $E=5$ direct comparisons: $T_1T_2$, $T_1T_3$, $T_1T_4$, $T_2T_3$, and $T_3T_4$. For this toy example, we specify summary effect estimates $\hat{\theta}^{T_1T_3}=2$, $\hat{\theta}^{T_1T_2} = \hat{\theta}^{T_2T_3} = 0.5$, and  $\hat{\theta}^{T_1T_4} = \hat{\theta}^{T_4T_3} = 1.5$. For our purposes, the choice of effect measure, whether it be risk ratio, mean difference, or any other metric, is not important. We assume all the studies have only two treatments (arms), and the variances of all summary effects are equal. For the latter, we choose an arbitrary value of $\text{var}^{T_iT_j} = 0.3^2$.

In the following, we use this toy example in two ways. First, we compare inconsistency measures for the treatment contrast $T_1$ vs $T_3$. For this comparison, we have a direct estimate $\hat{\theta}^{T_1T_3}=2$ and two indirect estimates, one via treatment $T_2$ ($T_1 \to T_2 \to T_3$), and one via treatment  $T_4$ ($T_1 \to T_4 \to T_3$). In Figure \ref{fig:fig_2}(b), these paths are shown in orange and red respectively. The corresponding estimates are $\hat{\theta}^{T_1T_3\text{(indir)}}_1 = \hat{\theta}^{T_1T_2} + \hat{\theta}^{T_2T_3} = 1$ and $\hat{\theta}^{T_1T_3\text{(indir)}}_2 = \hat{\theta}^{T_1T_4} + \hat{\theta}^{T_4T_3} = 3$. In this example, the three estimates of contrast $T_1T_3$ disagree, indicating some level of inconsistency. However, pooling the indirect estimates (with equal weights) leads to an overall indirect estimate $\hat{\theta}^{T_1T_3\text{(indir)}}=\frac{1}{2}(\hat{\theta}^{T_1T_3\text{(indir)}}_1 + \hat{\theta}^{T_1T_3\text{(indir)}}_2)=2$ which is equal to the direct estimate. Therefore, we use this example to illustrate how the methods perform when inconsistencies between different indirect estimates are masked by the pooled indirect evidence. Next, we focus on the contrast $T_2$ vs $T_3$ shown in Figure \ref{fig:fig_2}(c). As before, we have one direct estimate and two indirect estimates; the first via treatment $T_1$ ($T_2 \to T_1 \to T_3$) and the second via $T_1$ and $T_4$ ($T_2 \to T_1 \to T_4 \to T_3$). Both indirect estimates involve the edge $T_2T_1$. Therefore, we use this example to demonstrate how our method acts when different sources of evidence overlap.

\subsection{Inconsistency Assessments}

\subsubsection{Loop-specific} The loop-specific method tests for local inconsistencies by examining closed loops of evidence in the network created by multiple direct comparisons.\cite{Bucher:singleLoop} When these loops are independent, we test for differences between direct and indirect estimates in each loop. For a given loop $\ell_l$, we perform a Z-test on the difference, 
\begin{equation}
    \omega_{\ell_{l}} = |\theta^{T_iT_j\text{(dir)}} - \theta^{T_iT_j\text{(indir)}}_{\ell_{l}}|
    \label{eq:omega},
\end{equation}
where the null hypothesis assumes no inconsistency within the loop. The results from this test provide a p-value that indicates the likelihood of inconsistency.

For the $T_1T_3$ comparison in our toy example, there are two independent loops: $\ell_1 = T_1T_2T_3$ and $\ell_2 = T_1T_4T_3$. Following the loop-specific approach, one gets the same results for  $\omega_{\ell_1}$ and $\omega_{\ell_2}$. This is because the discrepancies between the direct estimate and each of the indirect estimates are the same,
\begin{align*}
        \omega_{\ell_1} &= |\hat{\theta}^{T_1T_3\text{(dir)}} - \hat{\theta}^{T_1T_3\text{(indir)}}_{\ell_1}|= 2 - 1 = 1,  \\
        \omega_{\ell_2} &= |\hat{\theta}^{T_1T_3\text{(dir)}} - \hat{\theta}^{T_1T_3\text{(indir)}}_{\ell_2}|= 3 - 2 = 1.
\end{align*}
The p-value for both loops is $0.05$ (see Table \ref{tab:pValues}). With a $95\%$ significance level, the loop-specific method correctly detects the inconsistency on $T_1T_3$. However, this approach can become complicated as the number of loops increases, making it difficult to identify independent loops within larger networks, which can lead to multiple testing. \cite{Higgings:design-by-treatment}

When the contrast of interest is $T_2$ vs $T_3$, the loops are $\ell_1 = T_2T_1T_3$ and $\ell_2 = T_2T_1T_4T_3$. These loops do not have the same symmetry as the first case, leading to two different p-values for each loop: $\text{p-value}_{\ell_1}=0.05$ and $\text{p-value}_{\ell_2}=0.0008$.

\subsubsection{Side-splitting} Rather than focusing on closed loops of evidence, the side-splitting method compares the different sources of evidence on each treatment comparison.\cite{Dias:book, Ian:side-splittingName} Here, the direct evidence is compared to the pooled indirect estimate using the inconsistency factor, $\omega^{T_iT_j}$, where 
\[
\omega^{T_iT_j} = |\theta^{T_iT_j\text{(dir)}} - \theta^{T_iT_j\text{(indir)}}_{total}|.
\]
A Z-test is performed to assess the statistical significance of the inconsistency.\cite{Dias:book} By lumping together the indirect evidence, this approach can obscure inconsistencies between different indirect estimates. This is the case for the $T_1T_3$ comparison in our toy example, where the pooled indirect evidence equals the direct estimate, resulting in no detected inconsistency ($\hat{\omega}^{T_1T_3} = 0$, $\text{p-value}= 1$).  Therefore, the inconsistency between indirect estimates is masked when we focus on the $T_1T_3$ comparison because the total indirect estimate equals the direct one. Conversely, the side-splitting approach does detect inconsistency ($\text{p-value}= 0.006$) for the treatment contrast $T_2$ vs $T_3$. The inconsistency in the network manifests in the results for this comparison.

\subsubsection{Design-by-treatment interaction} The previous two methods are described as `local' as they test for inconsistencies in particular areas of the network. The design-by-treatment interaction model instead calculates a global measure of inconsistency, which attempts to capture the inconsistency in the network as a whole. By introducing a parameter that captures differences between direct and indirect evidence, this method describes consistency and inconsistency models as multivariate meta-regressions.\cite{Higgings:design-by-treatment} Using a Wald statistic test on the inconsistency parameter, the result for our toy example, shown in Table \ref{tab:pValues}, indicates a p-value of $0.003$.  With a $95\%$ significance level, this suggests that there is global inconsistency within the network. However, while the model successfully identifies the presence of inconsistency, it does not provide information about where the inconsistency is located. 

\begin{center}
\begin{table*}[!h]%
\caption{Comparing different methods for detecting the inconsistency for $T_1T_3$ and $T_2T_3$ comparisons. Design-by-treatment gives a global measure of inconsistency in the whole network.\label{tab:pValues}}
\begin{tabular*}{\textwidth}{@{\extracolsep\fill}lllll@{}}
\toprule
&\multicolumn{1}{@{}l}{\textbf{Loop-specific}} &\multicolumn{1}{@{}l}{\textbf{Side-splitting}} &\multicolumn{1}{@{}l}{\textbf{Path-based}} & \multicolumn{1}{@{}l}{\textbf{Design-by-treatment}} \\\cmidrule{2-2}\cmidrule{3-3}\cmidrule{4-4}\cmidrule{5-5}
\textbf{Measure of (in)consistency} & \textbf{$\omega_{\ell_i}$}  & {\textbf{$\omega^{T_iT_j}$}} & \textbf{$Q^{path}_{T_iT_j}$} & \textbf{$Q^{DBT}$}   \\
\midrule
p-value($T_1T_3$) & $\ell_{(T_1T_2T_3)} \equiv \ell_{(T_1T_4T_3)}$: 0.05  & 1  & 0.003  &  -   \\
p-value($T_2T_3$) & $\ell_{(T_2T_1T_3)}$: 0.05  & 0.006  & 0.003 & -   \\
                  & $\ell_{(T_2T_1T_4T_3)}$:0.0008  &  - &  - &  -  \\
p-value($global$) &  - &  - & - & 0.003   \\
\bottomrule
\end{tabular*}
\end{table*}
\end{center}

\section{Path-Based Approach} \label{sec3}
\subsection{Definition of the Approach}
In Section \ref{sec2}, we saw that the side-splitting method could not properly capture the conflicting evidence on the $T_1T_3$ contrast in our toy example in Figure \ref{fig:fig_2}(b). The loop-specific method, while more accurate, measures the inconsistency of an entire loop instead of a single comparison, and its implementation in large networks may prove to be resource-intensive and lead to multiple testing. On the other hand, the design-by-treatment interaction model is a global method, and does not provide a local analysis of the inconsistency. It also relies on modeling assumptions (for instance, linearity between the treatment effects and the interaction term), and results may be sensitive to the chosen statistical model. If the model assumptions are not met, the results may be biased or misleading.\cite{Higgings:design-by-treatment}

To address these limitations, we propose a new measure of inconsistency based on "paths of evidence".\cite{Theo:path} Previous applications of paths  have demonstrated their utility in calculating the proportion contributions of direct comparisons in the context of NMA estimates.\cite{Krahn:path, Theo:path, Anna:RW, Gerta:shortest-path} Specifically, graph theory has been employed to derive these proportions through the use of the hat matrix ($\textbf{\text{H}}$), which translates into directed networks.\cite{Konig_visualise_flow, Theo:path, Anna:RW} Here, a path refers to a series of distinct nodes connected by direct comparisons. By using a path-based approach, we ensure that we do not overlook conflicts that may arise from lumping all the indirect evidence. Instead, we consider each piece of indirect evidence individually. An additional benefit of this approach is that it does not require direct evidence to draw inferences regarding the inconsistency between two treatments. Even in the absence of direct evidence, we can still investigate the consistency between indirect paths, which represent different sources of evidence.

Using the hat matrix, $\textbf{\text{H}}$, we start with the directed network for two treatments of interest, $T_i$ and $T_j$. A path $\pi^{T_iT_j}_p$, is a series of  distinct  edges (or nodes) that starts at node $T_i$, passes through edges in the direction given by the sign of the corresponding entry in the $\textbf{\text{H}}$ matrix, and ends at node $T_j$. In another words, we can write a path as a set of distinct edges  $\pi^{T_iT_j}_p = \{T_iT_k, \dots, T_{k^\prime}T_j\}$.
The total number of paths between the pair of nodes $T_i$ and $T_j$ is denoted by $P$, with $\pi^{T_iT_j}$ representing the complete set of such paths, $\pi^{T_iT_j} = \{\pi^{T_iT_j}_1, \dots, \pi^{T_iT_j}_{P}\}$. In this context, the direct comparison between $T_i$ and $T_j$ is just one path out of $P$ paths between these two treatments, with only one edge in the path.

Each path is associated with an effect, $\theta(\pi_p^{T_iT_j})$, and a variance, $\text{var}(\pi_p^{T_iT_j})$. The path effect is equal to the summation of the effects in the edges involved in that path,
\[
\theta(\pi_p^{T_iT_j}) = \sum_{T_kT_{k^\prime}\in \pi^{T_iT_j}_p} \theta^{T_kT_{k^\prime}}.
\]
Similarly, the path variance is the summation of the corresponding variances,
\[
\text{var}(\pi_p^{T_iT_j}) = \sum_{T_kT_{k^\prime}\in \pi^{T_iT_j}_p} \text{var}^{T_kT_{k^\prime}},
\] 
where $\text{var}^{T_kT_{k^\prime}}$ is the variance for the the comparison $T_kT_{k^\prime}$. 

\begin{figure*}
    \centering
    \includegraphics[width=0.8\textwidth]{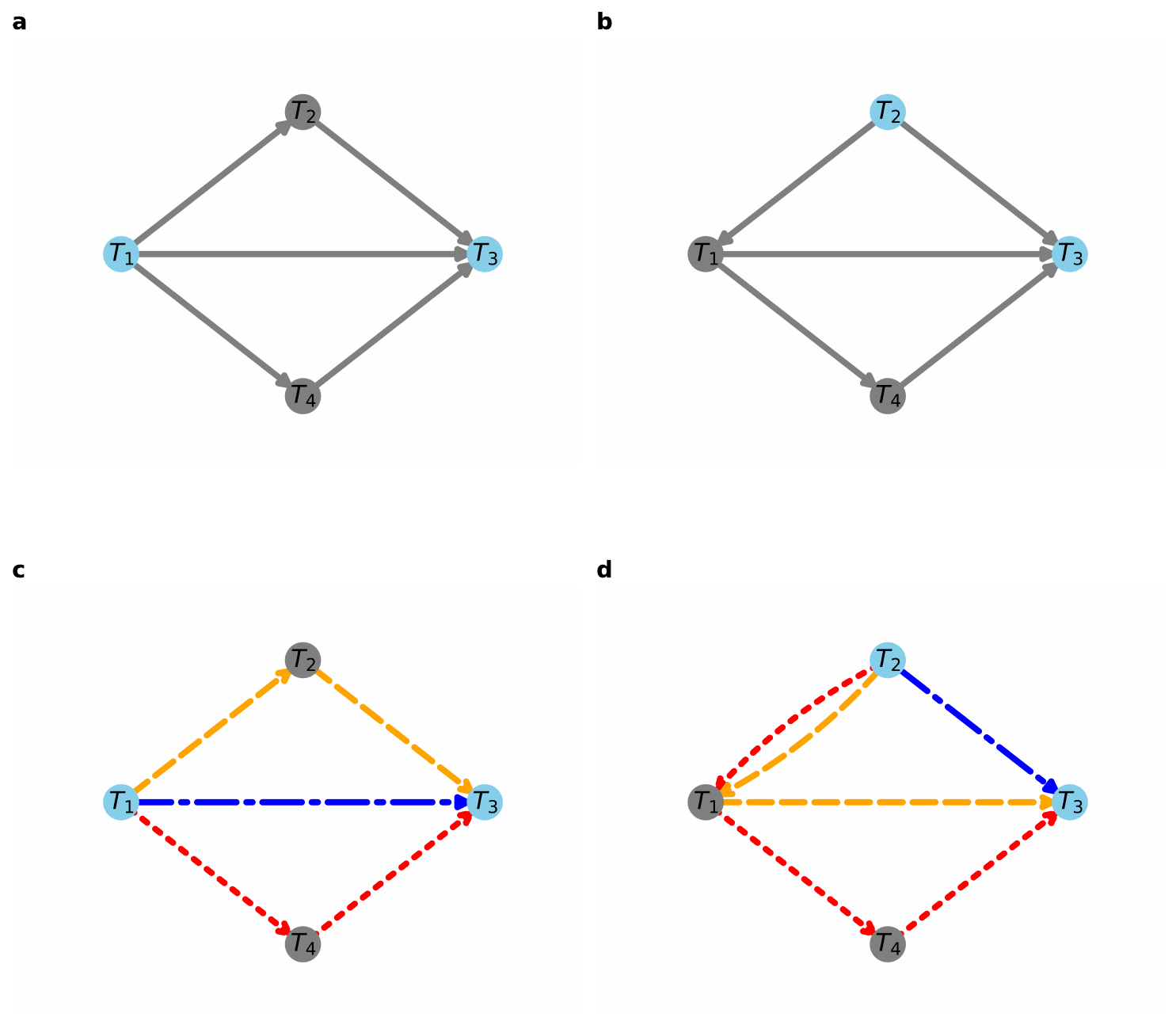}
    \caption{The directed  network constructed from the $\textbf{\text{H}}$ matrix for (a) $T_1$ vs $T_3$ and (b) $T_2$ vs $T_3$. All paths of evidence from (c) $T_1$ to $T_3$ and (d) $T_2$ to $T_3$. Direct evidence is shown in dot-dashed blue. Indirect paths are shown in dashed orange and dotted red.}
    \label{fig:fig_3}
\end{figure*}

Figures \ref{fig:fig_3}(a) and \ref{fig:fig_3}(b) show the directed networks for comparisons $T_1:T_3$ and $T_2:T_3$ in our fictional example.
As shown in Figures \ref{fig:fig_3}(c) and \ref{fig:fig_3}(d), each of these comparisons has three paths connecting the treatments of interest,
\begin{align*}
    &\begin{aligned}
        T_iT_j & = T_1T_3 \\
        \pi_1^{T_1T_3} &: T_1 \rightarrow  T_3\\
        \pi_2^{T_1T_3} &: T_1 \rightarrow T_2 \rightarrow  T_3\\
        \pi_3^{T_1T_3} &: T_1 \rightarrow T_4 \rightarrow  T_3,
    \end{aligned}
    &&
    \begin{aligned}
        T_iT_j & = T_2T_3 \\
        \pi_1^{T_2T_3} &: T_2 \rightarrow  T_3\\
        \pi_2^{T_2T_3} &: T_2 \rightarrow T_1 \rightarrow  T_3\\
        \pi_3^{T_2T_3} &: T_2 \rightarrow T_1 \rightarrow T_4 \rightarrow  T_3.
    \end{aligned}
\end{align*}
For each comparison of interest, we construct two vectors: (1) $\boldsymbol{\Theta}(\pi^{T_iT_j}) = (\theta(\pi^{T_iT_j}_1), \dots, \theta(\pi^{T_iT_j}_{P}))^\top$, a vector of length  $P$ containing the effect estimates of each path, and (2) $\hat{\boldsymbol{\Theta}}^{{\text{$T_iT_j$(nma)}}}=\hat{\theta}^{{\text{$T_iT_j$(nma)}}} I_{P\times1}$ containing the network estimate of that comparison. For our fictional example these vectors are 
\begin{align*}
\begin{aligned}
            \hat{\boldsymbol{\Theta}}(\pi^{T_1T_3}) &= \begin{pmatrix}
            \hat{\theta}(\pi^{T_1T_3}_1)\\
            \hat{\theta}(\pi^{T_1T_3}_2)\\
            \hat{\theta}(\pi^{T_1T_3}_3)
        \end{pmatrix} &= \begin{pmatrix}
            2\\
            1\\
            3
        \end{pmatrix} \\
         \hat{\boldsymbol{\Theta}}^{T_1T_3\text{(nma)}} &= \begin{pmatrix}
            \hat{\theta}^{T_1T_3\text{(nma)}}\\
            \hat{\theta}^{T_1T_3\text{(nma)}}\\
            \hat{\theta}^{T_1T_3\text{(nma)}}
        \end{pmatrix} &= \begin{pmatrix}
            2\\
            2\\
            2
        \end{pmatrix}
\end{aligned}
&&
\begin{aligned}
            \hat{\boldsymbol{\Theta}}(\pi^{T_2T_3}) &= \begin{pmatrix}
            \hat{\theta}(\pi^{T_2T_3}_1)\\
            \hat{\theta}(\pi^{T_2T_3}_2)\\
            \hat{\theta}(\pi^{T_2T_3}_3)
        \end{pmatrix} &= \begin{pmatrix}
            0.5\\
            1.5\\
            2.5
        \end{pmatrix}\\
        \hat{\boldsymbol{\Theta}}^{T_2T_3\text{(nma)}} &= \begin{pmatrix}
            \hat{\theta}^{T_2T_3\text{(nma)}}\\
            \hat{\theta}^{T_2T_3\text{(nma)}}\\
            \hat{\theta}^{T_2T_3\text{(nma)}}
        \end{pmatrix} &= \begin{pmatrix}
            1\\
            1\\
            1
        \end{pmatrix}.
\end{aligned}
\end{align*}

We define $\boldsymbol{\Sigma}^{T_iT_j}$ as a $P \times P$ symmetric variance-covariance matrix where each row and column represents a path. The diagonal elements of this matrix 
are the variances of each path, and the off-diagonal elements 
are the covariance between each pair of paths,
\begin{equation}
\label{eq:cov}
    \boldsymbol{\Sigma}^{T_iT_j} = \begin{pmatrix}
        \text{var}(\pi_1^{T_iT_j}) &\text{cov}(\pi_1^{T_iT_j}, \pi_2^{T_iT_j}) &\dots &\text{cov}(\pi_1^{T_iT_j}, \pi_{P}^{T_iT_j})\\
        \text{cov}(\pi_2^{T_iT_j}, \pi_1^{T_iT_j}) &\text{var}(\pi_2^{T_iT_j}) &\dots &\text{cov}(\pi_2^{T_iT_j}, \pi_{P}^{T_iT_j})\\
        \vdots\\
        \text{cov}(\pi_{P}^{T_iT_j}, \pi_1^{T_iT_j}) &\dots &\dots &\text{var}(\pi_{P}^{T_iT_j})
    \end{pmatrix}.
\end{equation}
The covariance between two paths is the summation of the variances of all the edges that the two paths have in common. Assuming all edges in our fictional example have the same variance ($0.3^2$), we find for $T_iT_j = T_1T_3$ 
\begin{align}\label{eq:covAC}
    \boldsymbol{\Sigma}^{T_1T_3} &= \begin{pmatrix}
        \text{var}^{T_1T_3} &0 &0\\
        0 &\text{var}^{T_1T_3} + \text{var}^{T_2T_3} &0\\
        0 &0 &\text{var}^{T_1T_3} + \text{var}^{T_4T_3} 
    \end{pmatrix}  &= \begin{pmatrix}
        0.3^2 &0 &0\\
        0 &2(0.3^2) &0\\
        0 &0 &2(0.3^2)
    \end{pmatrix}, 
\end{align}
and for $T_iT_j = T_2T_3$, 
\begin{align}\label{eq:covBC}
    \boldsymbol{\Sigma}^{T_2T_3} &= \begin{pmatrix}
        \text{var}^{T_2T_3} &0 &0\\
        0 &\text{var}^{T_2T_1} + \text{var}^{T_1T_3} &\text{var}^{T_2T_1}\\
        0 &\text{var}^{T_2T_1} &\text{var}^{T_2T_1} + \text{var}^{T_1T_4} + \text{var}^{T_4T_3}
    \end{pmatrix} &= \begin{pmatrix}
        0.3^2 &0 &0\\
        0 &2(0.3^2) &0.3^2\\
        0 &0.3^2 &3(0.3^2)
    \end{pmatrix}.
\end{align}

One can show that the network estimate for a pair of treatments is equal to a weighted mean of normally distributed path estimates for that comparison.\cite{Gerta:shortest-path} We define a statistic $Q^{\text{path}}_{T_iT_j}$ to measure the inconsistency between different paths of evidence for the comparison $T_iT_j$,
\begin{equation}
    Q^{\text{path}}_{T_iT_j} = (\boldsymbol{\Theta}(\pi^{T_iT_j}) - \hat{\boldsymbol{\Theta}}^{T_iT_j(\text{nma})})^{\top}(\boldsymbol{\Sigma}^{T_iT_j})^{-1} (\boldsymbol{\Theta}(\pi^{T_iT_j}) - \hat{\boldsymbol{\Theta}}^{T_iT_j(\text{nma})})
    \label{eq:QPath}.
\end{equation} 
For a normally distributed $n$-dimensional vector $\textbf{x} \sim N_n(\mu, \boldsymbol{\Sigma})$, with mean $\mu$ and covariance matrix $\boldsymbol{\Sigma}$, the expression $($\textbf{x}$ - \mu)^T (\boldsymbol{\Sigma})^{-1} ($\textbf{x}$ - \mu)$ follows a chi-squared distribution $\chi_n^2$ with $n$ degrees of freedom.\cite{Johnson:bookMultivariate}  In Equation \eqref{eq:QPath},  $\boldsymbol{\Theta}(\pi^{T_iT_j})$ represents the estimated path effects for two selected treatments $T_i$ and $T_j$. Since  $\hat{\boldsymbol{\Theta}}(\pi^{T_iT_j})$ and  $\hat{\boldsymbol{\Theta}}^{{T_iT_j}(\text{nma})}$ are estimated values based on observed variables, $Q^{\text{path}}_{T_iT_j}$ follows a chi-squared distribution  $\chi_{P-1}^2$ with $P-1$ degrees of freedom. Therefore, we can use this measure to obtain a p-value to test our null hypothesis of consistency. 

With a p-value of $0.003$ in Table \ref{tab:pValues}, this path-based approach effectively detects inconsistency in both the $T_1T_3$ and $T_2T_3$ comparisons in our toy example, even when it is masked in the $T_1T_3$ case.

\subsection{Linearly Dependent Paths} \label{sec:Complex_Paths}

\begin{figure*}
    \centering
    \includegraphics[width=0.8\textwidth]{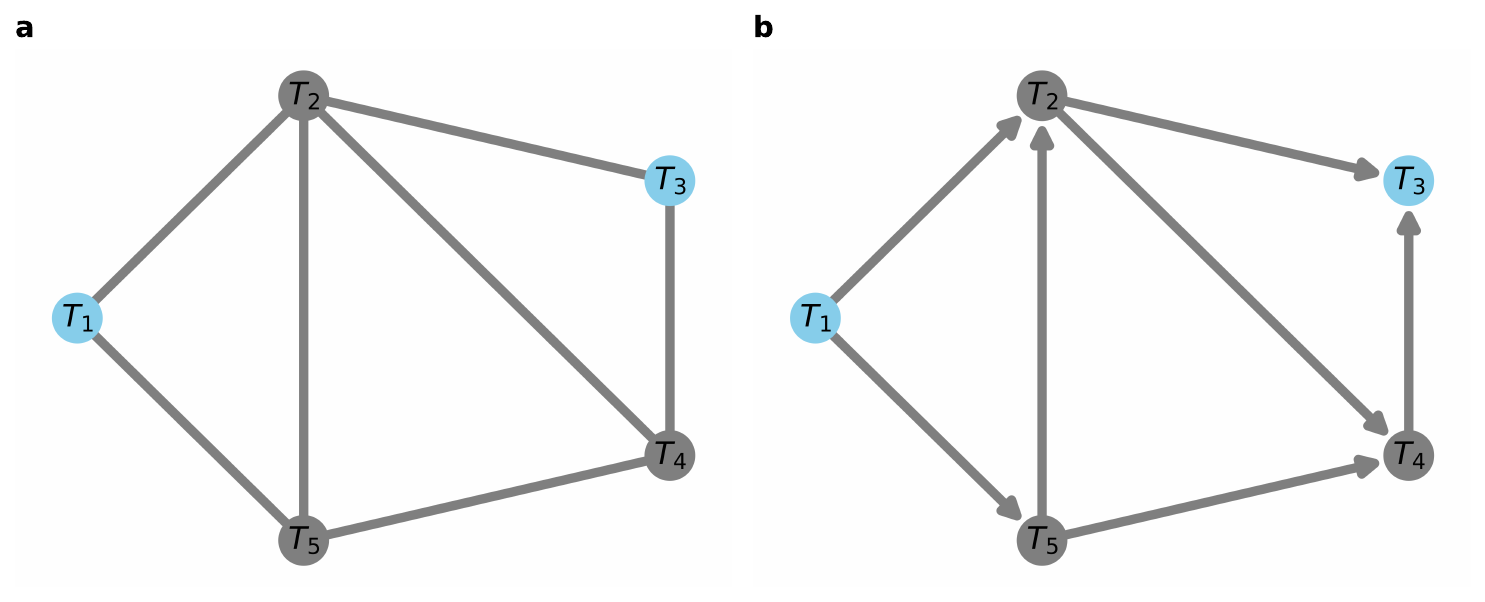}
    \caption{(a) Fictional example network from R{\"u}cker et al.\cite{Gerta:shortest-path} with five nodes and seven edges. (b) The directed network for the comparison $T_1$ vs $T_3$ constructed from the corresponding row of the $\textbf{\text{H}}$ matrix.}
    \label{fig:fig_4}
\end{figure*}

\begin{figure*}
    \centering
    \includegraphics[width=\textwidth]{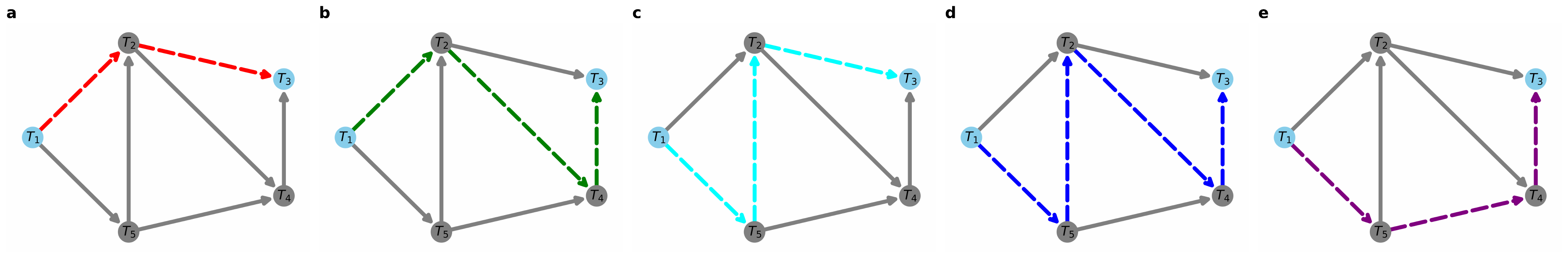}
    \caption{All the five paths of evidence between $T_1$ and $T_3$. (a) $\pi_1^{T_1T_3}$, (b) $\pi_2^{T_1T_3}$, (c) $\pi_3^{T_1T_3}$, (d) $\pi_4^{T_1T_3}$, and (e) $\pi_5^{T_1T_3}$.\cite{Gerta:shortest-path}}
    \label{fig:fig_5}
\end{figure*}

The path inconsistency statistic in Equation \eqref{eq:QPath}
can only be evaluated if the path covariance matrix $\boldsymbol{\Sigma}^{T_iT_j}$ is invertible. Since $\boldsymbol{\Sigma}^{T_iT_j}$ is square, a lack of inverse indicates the matrix is singular. This is caused when some paths are linear combinations of other paths. To illustrate this, we introduce a second fictional example from R{\"u}cker et al.\cite{Gerta:shortest-path} shown in Figure \ref{fig:fig_4}. This network has $N=5$ treatments and $E=7$ direct comparisons. We focus on the comparison of $T_1$ vs $T_3$. Figure \ref{fig:fig_4}(b) shows the directed network of evidence obtained from the corresponding row of the $\textbf{\text{H}}$ matrix. Due to a lack of direct evidence on this comparison, there is no direct path from $T_1$ to $T_3$. There are, however, five indirect paths labeled $\pi^{T_1T_3}_{1}$, $\dots$, $\pi^{T_1T_3}_{5}$ and shown in Figure \ref{fig:fig_5}.

\begin{figure*}
    \centering
    \includegraphics[width=0.8\textwidth]{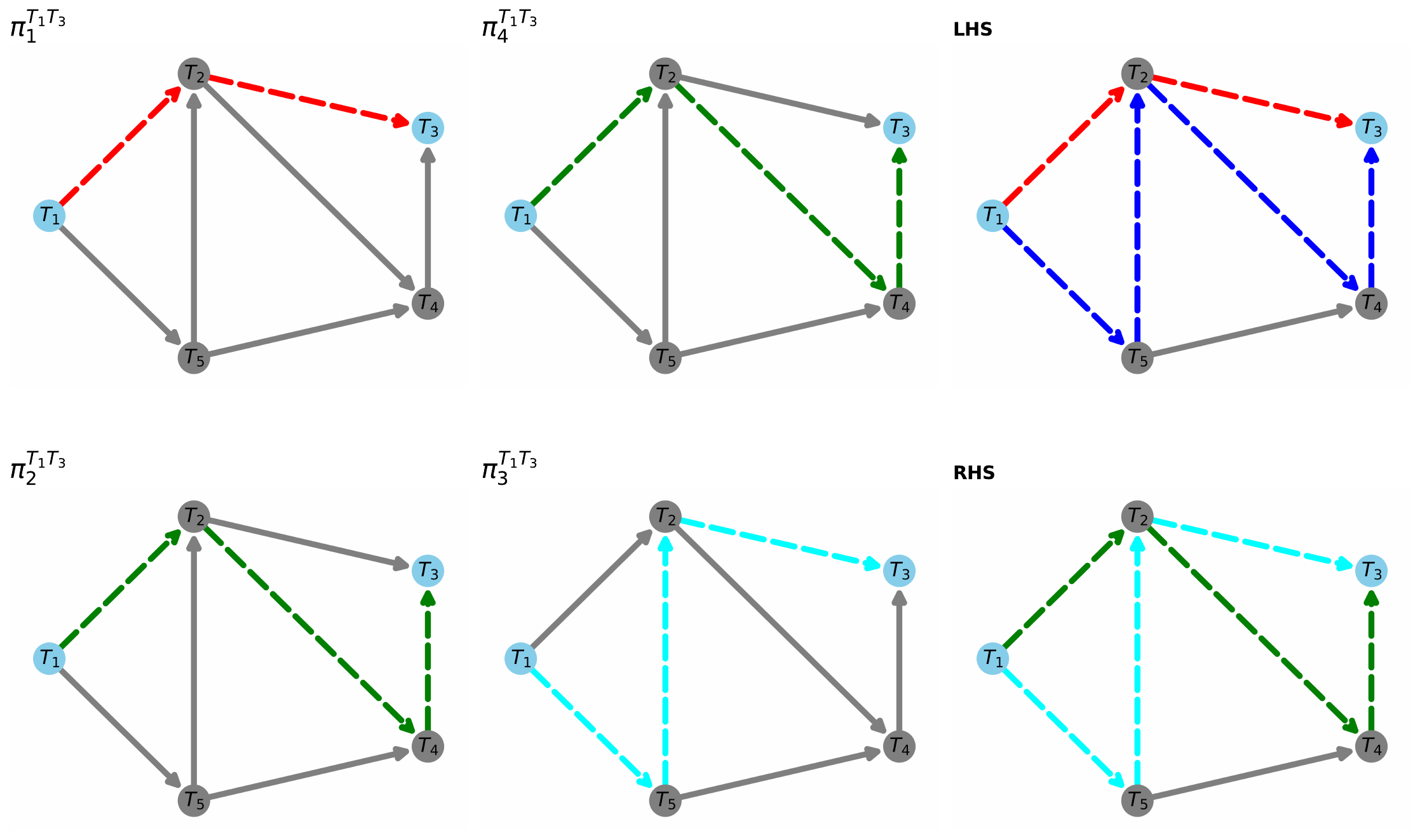}
    \caption{Depiction of the linear relation where the left-hand side(LHS), $\pi^{T_1T_3}_{1} + \pi^{T_1T_3}_{4}$, equals the right-hand side(RHS), $ \pi^{T_1T_3}_{2} + \pi^{T_1T_3}_{3}$.}
    \label{fig:fig_6}
\end{figure*}

We introduce the terms `adding' and `subtracting' paths to mean combining or removing all the edges involved in those paths. In Figure \ref{fig:fig_6}, we show that adding paths $\pi_1^{T_1T_3}$ and $\pi_4^{T_1T_3}$ gives the same result (the same set of directed edges) as adding paths $\pi_2^{T_1T_3}$ and $\pi_3^{T_1T_3}$. Therefore, we can write the linear relation \cite{Gerta:shortest-path}
\begin{equation}
    \pi^{T_1T_3}_{1} + \pi^{T_1T_3}_{4} = \pi^{T_1T_3}_{2} + \pi^{T_1T_3}_{3}.
    \label{eq:linCombo}
\end{equation}
We call these paths "linearly dependent" \cite{Gerta:shortest-path} since any of the four paths $ \pi^{T_1T_3}_{1}$, $\pi^{T_1T_3}_{2}$, $\pi^{T_1T_3}_{3}$, and $\pi^{T_1T_3}_{4}$ can be created by adding and subtracting the other three according to Equation (\ref{eq:linCombo}). In general, a network may involve multiple linear relations between different subset of paths, especially if the network is very large. 

In Section \ref{app1.1a} of the Appendix, we show that the covariance matrix $\boldsymbol{\Sigma}^{T_iT_j}$ becomes invertible once all linear dependencies between paths are removed. Therefore, for each linearly dependent subset of paths in a network, we must break the dependency by removing the dependent paths. The choice of which path to remove does not affect the overall computation of $Q_{T_iT_j}^{\text{path}}$ because the respective path is removed from both the path estimate vector and the covariance matrix. Therefore, this choice is arbitrary. Detailed explanation via an example is given in \ref{app1.3a} of the Appendix. We write $P^{\prime}<P$ for the number of independent paths remaining once the dependent paths have been removed, and these are the paths that we use in Equation \ref{eq:QPath}.

To identify linearly dependent paths, we define $\textbf{\text{A}}^{T_iT_j}$ as a $P \times P$ symmetric path-adjacency matrix, where each row and column represents a path. Diagonal elements, $a_{dd}$, contain the length of the corresponding path $\pi^{T_iT_j}_{d}$ and off-diagonal elements, $a_{rc}$, are given by the number of shared edges between $\pi^{T_iT_j}_{r}$ and $\pi^{T_iT_j}_{c}$. $\textbf{\text{A}}^{T_iT_j}$ has full rank if and only if there is no dependency between paths. Therefore, we obtain a set of independent paths by reducing this matrix to its rank. For details, we refer to Section \ref{app1.2a} in the Appendix. Once the set of independent paths are identified, we define a reduced $P^{\prime} \times P^{\prime}$ path-covariance matrix $\boldsymbol{\Sigma}^{T_iT_j}$ via Equation \eqref{eq:cov} and a corresponding $P^{\prime} \times 1$ vector of path effects, $\boldsymbol{\Theta}(\pi^{T_iT_j})$.

\subsection{Netpath Plot, Beyond Statistical Testing}

For a particular treatment comparison $T_iT_j$, the path inconsistency method measures how far away the network estimate of that comparison is from the estimate associated with each path of evidence. These deviations are aggregated into a single statistic $Q_{T_iT_j}^{\text{path}}$ which is used to test for the presence of inconsistency on that comparison. Following this assessment, one might then ask the following questions: (1) which paths contribute to the inconsistency, and (2) how large is the discrepancy between their estimates? To answer these questions, we propose a method for visualizing the extent of disagreement between pairs of paths. This takes our approach beyond just a test for inconsistency, allowing for a more granular exploration of where and to what extent inconsistencies occur. 

\begin{figure*}
    \centering
    \includegraphics[width=\textwidth]{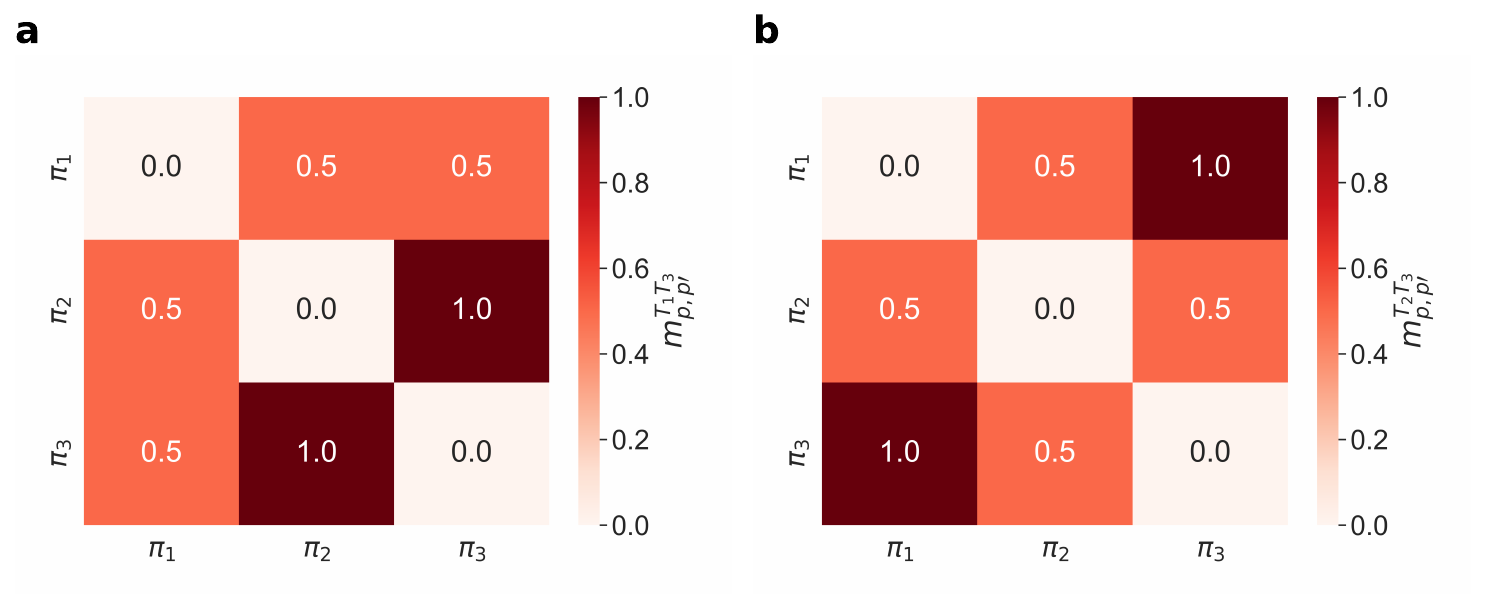}
    \caption{The Netpath plot to visualize Equation (\ref{eq:m}) for both cases of our first fictional example in Figures (a) \ref{fig:fig_3}(c) and (b) \ref{fig:fig_3}(d). The superscripts  in $\pi_i^{T_1T_3}$ and  $\pi_i^{T_2T_3}$ are omitted for a less cluttered representation.}
    \label{fig:fig_7}
\end{figure*}

To this aim, we define $m_{pp^{\prime}}^{T_iT_j}$ as the scaled difference between the path effects of paths $p$ and $p^{\prime}$,
\begin{equation}
    m_{p,p^{\prime}}^{T_iT_j} = \frac{|\theta(\pi^{T_iT_j}_p) - \theta(\pi^{T_iT_j}_{p^{\prime}})|}{max_{kl}(|\theta(\pi^{T_iT_j}_k) - \theta(\pi^{T_iT_j}_l)|)},
\label{eq:m}
\end{equation}
where the maximum in the denominator is over all pairs of paths $(\pi^{T_iT_j}_k, \pi^{T_iT_j}_l)$ between $T_i$ and $T_j$. Scaling by the largest difference in path effects ensures that the values for $m_{pp^{\prime}}^{T_iT_j}$  lie between $0$ and $1$. We plot these values as a heat map that shows the discrepancies between each pair of paths for a comparison $T_iT_j$. We call it Netpath plot. Panels (a) and (b) in Figure \ref{fig:fig_7} show this visualization  for comparisons $T_1T_3$ and $T_2T_3$, respectively, in our first toy example.  The darker colors indicate more conflict between the corresponding pair of paths. 

Figure \ref{fig:fig_7}(a) shows that the two indirect paths, $\pi^{T_1T_3}_2$ ($T_1 \to T_2 \to T_3$) and $\pi^{T_1T_3}_3$ ($T_1 \to T_4 \to T_3$), are equally inconsistent with the direct evidence $\pi^{T_1T_3}_1$ ($T_1 \to T_3$). This means that the (absolute) difference between the estimates associated with paths $\pi^{T_1T_3}_1$ and $\pi^{T_1T_3}_2$ is the same as the (absolute) difference between paths $\pi^{T_1T_3}_1$ and $\pi^{T_1T_3}_3$. Indeed, this is the symmetry we specified in our set-up that causes inconsistency to be masked when the indirect evidence is pooled\footnote{Note that this masking effect can still occur even without symmetry; we could have increased the discrepancy of one path and decreased the discrepancy on the other as long as their average was still equal to the direct estimate.}. As expected, the highest degree of inconsistency on $T_1T_3$ comparison is between the two indirect paths $\pi^{T_1T_3}_2$ and $\pi^{T_1T_3}_3$. 

Figure \ref{fig:fig_7}(b) is the Netpath plot corresponding to our second case ($T_iT_j  = T_2T_3$) in Figure \ref{fig:fig_3}(d). This plot displays that the direct path $\pi^{T_2T_3}_1$ ($T_2 \to T_3$) is in more conflict with $\pi^{T_2T_3}_3$ ($T_2 \to T_1 \to T_4 \to T_3$) than it is with  $\pi^{T_2T_3}_2$ ($T_2 \to T_1 \to T_3$). This arises from the fact that the discrepancy between the direct path estimate ($\hat{\theta}(\pi^{T_2T_3}_1) = 0.5$) and the path estimate $\hat{\theta}(\pi^{T_2T_3}_3) = 2.5$ is larger compared to the discrepancy with the path estimate $\hat{\theta}(\pi^{T_2T_3}_2) = 1.5$.

To implement the path-based inconsistency methods described here, we have added a new function {\tt netpath()} to the {\tt netmeta}    package.\cite{netmeta_git} It takes a netmeta object as its input, and outputs a summary of the path inconsistencies on each treatment comparison.
This summary includes degrees of freedom, a p-value for the null hypothesis test, and the Netpath plots in Figure \ref{fig:fig_7}. Additional detail is presented in the Supporting Information.


\section{Real-World Example} \label{sec4}
\begin{figure*}
    \centering
    \includegraphics[width=\textwidth]{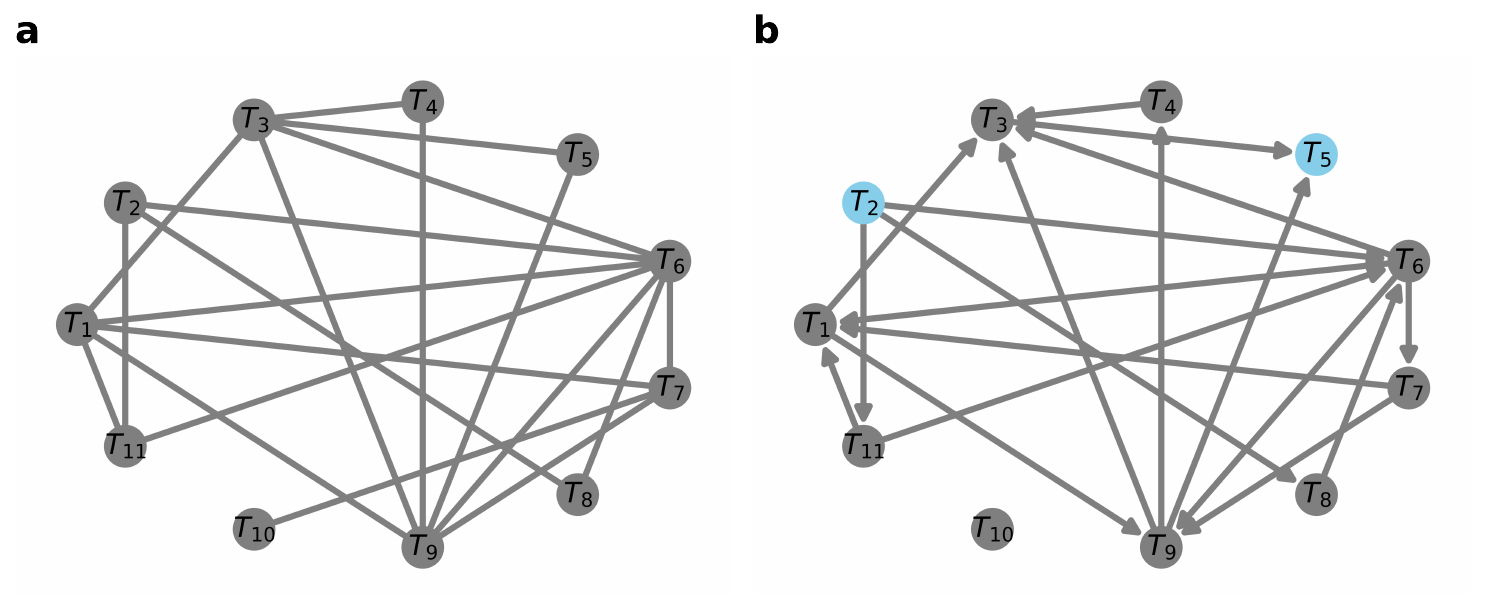}
    \caption{(a) The network graph for the NMA of depression.\cite{depressionData} This shows treatments $\{T_1, \dots, T_{11}\}$, with their clinical names provided in Footnote~\ref{node_names}. (b) The corresponding directed network for comparison $T_2:T_5$ constructed from the $\textbf{\text{H}}$ matrix.}
    \label{fig:fig_8}
\end{figure*}

Now, we apply our method to a real-world example. We choose an NMA of 11 treatments for depression,\cite{depressionData} where the primary outcome is a binary indicator of patient response after completing the treatment. We label the treatments by the set $\{T_1, \dots, T_{11}\}$\footnote{\label{node_names} $T_1$: tricyclic or tetracyclic antidepressants, $T_2$: selective serotonin reuptake inhibitors, $T_3$: psychotherapy + usual care, $T_4$: counselling + usual care, $T_5$: psycho-education + usual care, $T_6$: psychotherapy, $T_7$: counselling, $T_8$: psychotherapy + SSRIs, $T_9$: usual care, $T_{10}$: individualized antidepressant, $T_{11}$: placebo.}. The network includes one four-arm trial, eight three-arm, and 17 two-arm trials. There are a total of 20 direct comparisons (edges). The network graph is shown in Figure \ref{fig:fig_8}(a).

To demonstrate our approach, we focus on the comparison treatments $T_2$ vs $T_5$.  Although no direct evidence is available for this comparison, we can construct paths of evidence and assess the inconsistency between these two treatments. We start by calculating the $\textbf{\text{H}}$ matrix using {\tt hatmatrix()} function in {\tt netmeta}   package. Then, we construct a directed network of evidence for the two target treatments based on the $T_2:T_5$ row in the $\textbf{\text{H}}$ matrix. This network is shown in Figure \ref{fig:fig_8}(b).  We perform a depth-first search method \cite{DFS:book} on this directed network, uncovering a total of $49$ paths from \(T_2\) to \(T_5\). Consequently, the path-adjacency matrix for this comparison, \(\textbf{\text{A}}^{T_2T_5}\), is of dimensions \(49 \times 49\).  However, not all of these paths are linearly independent. Thus, we reduce the \(\textbf{\text{A}}^{T_2T_5}_{49 \times 49}\) matrix to its rank of \(11\), resulting in $P^{\prime} = 11$. These \(11\) independent paths are:
\begin{multicols}{2} 
\begin{itemize}
    \item $\pi_1 : \{T_2 \to T_6 \to T_1 \to T_3 \to T_5\}$
    \item $\pi_2 : \{T_2 \to T_6 \to T_1 \to T_9 \to T_3 \to T_5\}$
    \item $\pi_3 : \{T_2 \to T_6 \to T_1 \to T_9 \to T_4 \to T_3 \to T_5\}$
    \item $\pi_4 : \{T_2 \to T_6 \to T_1 \to T_9 \to T_5\}$
    \item $\pi_5 : \{T_2 \to T_6 \to T_3 \to T_5\}$
    \item $\pi_6 : \{T_2 \to T_6 \to T_7 \to T_1 \to T_3 \to T_5\}$
    \item $\pi_7 : \{T_2 \to T_6 \to T_7 \to T_9 \to T_3 \to T_5\}$
    \item $\pi_8 : \{T_2 \to T_6 \to T_9 \to T_3 \to T_5\}$
    \item $\pi_9 : \{T_2 \to T_8 \to T_6 \to T_1 \to T_3 \to T_5\}$
    \item $\pi_{10} : \{T_2 \to T_{11} \to T_1 \to T_3 \to T_5\}$
    \item $\pi_{11} : \{T_2 \to T_{11} \to T_6 \to T_1 \to T_3 \to T_5\}$
\end{itemize}
\end{multicols}
\noindent   where we have omitted the superscript in $\pi_i^{T_2T_5}$  for clarity. Statistical details such as the effect estimate and total variance for these paths is given in in Supporting Information.

These \(11\) independent paths enables us to construct the full-rank invertible variance-covariance matrix, $\boldsymbol{\Sigma}^{T_2T_5}_{11\times 11}$ which is presented in Supporting Information. Next, we calculate the paths estimate vector for the $11$ independent paths,
\[
\hat{\boldsymbol{\Theta}}(\pi^{T_2T_5}) = (
    1.37 , 1.04 , 1.13 , 0.87 , 
    1.27 , 1.50 , 0.44 , 0.56 , 
    1.39 , -0.06 , 0.95 )_{1 \times 11}, 
\]
 where each effect estimate is a log odds ratio. The network estimate in Equation \ref{eq:nma_estimate} is calculated as $\hat{\theta}^{T_2T_5\text{(nma)}} = 0.458$, which is obtained from a  common-effect model using the {\tt netmeta}   package.

\begin{figure*}
    \centering
    \includegraphics[width=\textwidth]{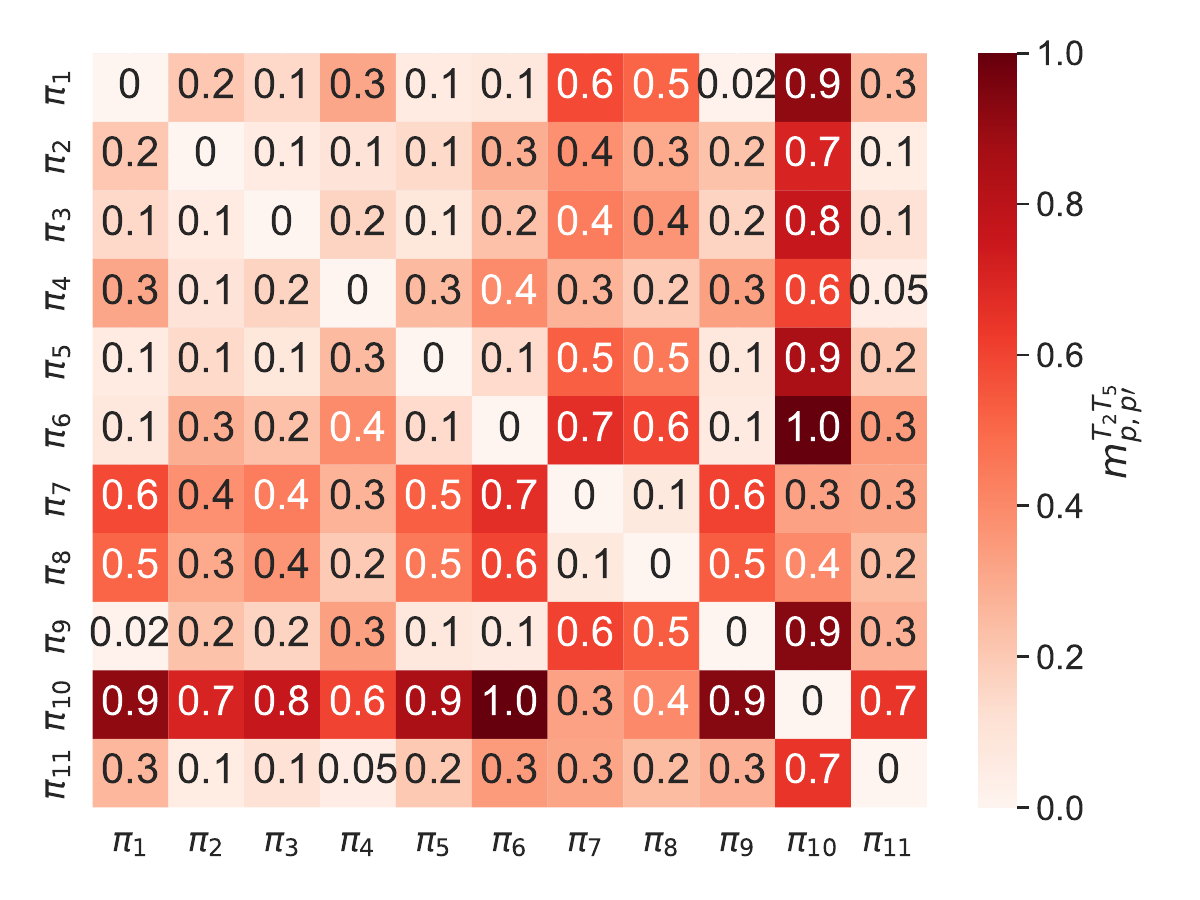}
    \caption{Inconsistency Netpath plot for all the independent paths between $T_2$ and $T_5$ in Figure \ref{fig:fig_8}(b). The superscript  in $\pi_i^{T_2T_5}$  is omitted for clarity.}
    \label{fig:fig_9}
\end{figure*}

Now, we insert  $\hat{\boldsymbol{\Theta}}(\pi^{T_2T_5})$, $\hat{\theta}^{T_2T_5\text{(nma)}}$, and $\boldsymbol{\Sigma}^{T_2T_5}$ into Equation \ref{eq:QPath}. We obtain $Q_{T_2T_5}^{\text{path}} = 9.451$; which, based on the assumption of  a $\chi^2_{11-1}$ distribution with $10$ degrees of freedom, results in a p-value of $0.49$. At the $95\%$ significance level, this result suggests there is no significant inconsistency on this comparison. However, this should not stop us from further investigation. We present Figure \ref{fig:fig_9} for more details on how consistent different paths of evidence are compared to one another.

The Netpath plot in Figure \ref{fig:fig_9} reveals that, among all pairs of paths, the greatest conflict is between the estimates associated with paths $\pi_6^{T_2T_5}$ and $\pi_{10}^{T_2T_5}$ where $m^{T_2T_5}_{6,10}= 1$. At the other extreme, paths $\pi_1^{T_2T_5}$ and $\pi_{9}^{T_2T_5}$ show the least disagreement, with the minimum value of $m^{T_2T_5}_{1,9}= 0.02$.  Paths $\pi_4^{T_2T_5}$ and $\pi_{11}^{T_2T_5}$ have the second least disagreement, associated with the value of  $m^{T_2T_5}_{4,11}= 0.05$. Further insights are also possible. For example,  we observe that $\pi_{10}^{^{T_2T_5}}$, overall, is in the most disagreement with the other paths while $\pi_{2}^{^{T_2T_5}}$ is the most similar to the others.

\section{Discussion}
In this paper, we address the issue of assessing inconsistency in networks of treatment comparisons. Traditional methods, which rely on the separation of direct from indirect evidence, have limitations when the direct comparison is unavailable. They also struggle in certain cases where the total indirect evidence aligns with the direct evidence. To overcome these challenges, we proposed a path-based approach. Additionally, we reviewed established methodologies for detecting and assessing inconsistency, highlighting their strengths and limitations.

One reason for the widespread use of NMA is its ability to address key questions that decision-makers face when considering multiple treatment options rather than just two.\cite{higgins:decision_making} As discussed, consistency is one of the underlying assumptions in NMA, and assessing it is crucial for the validity of the results. By adopting a path perspective, our proposed method, involving a novel inconsistency measure, has the potential to improve the assessment of inconsistency between multiple sources of evidence in  NMA. This contributes to more reliable results for end-users interested in specific treatment comparisons.

We investigate the complete network of evidence and introduce a quantitative measure of inconsistency for any comparison of interest that aims to identify discrepancies between different indirect paths of evidence. Our method evaluates each piece of evidence individually, avoiding the pitfalls of lumping together all indirect evidence. Furthermore, our approach does not rely  on the presence of direct evidence, enabling a thorough investigation of inconsistency for comparisons where the direct evidence is absent. Additionally, we go beyond merely testing for inconsistency; we illustrate the extent of inconsistencies between different sources of evidence using a Netpath plot. By adopting a path perspective, we can identify conflicts arising from varying effect estimations in a network of evidence.

We used two toy examples and one real-world dataset to underscore the practical applicability of our approach, showcasing its detail in managing  networks of treatment. We illustrated the step-by-step process of our method, highlighting how our path-based approach, combined with the new inconsistency measure, can identify and address inconsistencies. Our method, along with the proposed Netpath plot, provides a comprehensive examination of the entire network of evidence while also facilitating localized insights.

Our findings highlight the effectiveness of the proposed method in identifying inconsistency within a network of treatments. However, inconsistency is closely linked to the assumption of transitivity, which remains a fundamental yet complicated concept in NMA. Transitivity implies that if treatments are comparable through an intermediate common comparator, then indirect comparisons are valid. Ensuring that this assumption holds in practice is challenging due to potential differences in study characteristics or patient populations. If transitivity is violated, it can manifest as inconsistency within the network, but discerning the root cause of such inconsistency is not always straightforward. In this context, the observed inconsistencies may arise from either genuine violations of transitivity, model misspecifications, or random variability. As a result, current ambiguity of the relationship between transitivity and consistency can limit our ability to interpret and address transitivity fully. Therefore, while our method offers a valuable tool for exploring and quantifying inconsistency, careful attention to the transitivity assumption remains crucial for drawing reliable conclusions from NMA.

\section*{Financial Disclosure}
Authors Noosheen R. Tahmasebi, Theodoros Papakonstantinou, and Adriani Nikolakopoulou were supported by the Deutsche Forschungsgemeinschaft (DFG, German Research Foundation) - grant number NI 2226/1-1. Adriani Nikolakopoulou was also supported by DFG Project-ID 499552394 – SFB 1597. Author Annabel L Davies received funding from the Engineering and Physical Sciences Research Council (EPSRC) EP/Y007905/1.

\section*{Conflict of Interest}

The authors declare no potential conflict of interests.

\textbf{For further inquiries, please contact the author Noosheen R. Tahmasebi at noosheen.rajabzadehtahmasebi@uniklinik-freiburg.de}

\setcounter{section}{0}
\renewcommand{\thesection}{A\arabic{section}} 
\section*{Appendix \label{app1}}


\section{Singularity of the Covariance Matrix \label{app1.1a}}

In this part, we prove that  by removing the linearly dependent paths, the covariance matrix $\boldsymbol{\Sigma}^{T_iT_j}$ will be always invertible unless there is a direct comparison with a variance of $0$.

We define the path-edge incidence matrix $\text{\textbf{C}}^{T_iT_j}$ as a $P \times E$ matrix where each row represents an edge (direct comparison), and each column represents a path between $T_i$ and $T_j$ . In this matrix, an element is $1$ if the corresponding edge is part of the path, and $0$ otherwise. The path-adjacency matrix, $\text{\textbf{A}}^{T_iT_j}$, can then be written as 
\begin{equation}
    \text{\textbf{A}}^{T_iT_j} = \text{\textbf{C}}^{T_iT_j}.I.(\text{\textbf{C}}^{T_iT_j})^{\top},
    \label{eq:A}
\end{equation}
 where $I$ is an $E \times E$ identity matrix. Using the path-edge incidence matrix, we can also define the variance-covariance matrix as 
 \begin{equation}
     \boldsymbol{\Sigma}^{T_iT_j} = \text{\textbf{C}}^{T_iT_j}.\text{\textbf{V}}.(\text{\textbf{C}}^{T_iT_j})^{\top},
     \label{eq:Sig}
 \end{equation}
 where $\text{\textbf{V}}$ is an $E\times E$ diagonal matrix of edge variances, with diagonal elements $v_{dd} = \text{var}^{T_kT_{k^{\prime}}}$. 

The determinant of a diagonal matrix is  the product of its diagonal elements, which means $\text{det}(\bm{I}) = 1$ and $\text{det}(\text{\textbf{V}}) = \displaystyle \prod_{d=1}^{E} v_{dd}$. If $\text{\textbf{A}}^{T_iT_j}$ is not singular, $\text{\textbf{C}}^{T_iT_j}$ must have a non-zero determinant according to Equation \ref{eq:A}, therefore, $\boldsymbol{\Sigma}^{T_iT_j}$ cannot be singular unless the variance of a direct comparison $v_{dd}$ is $0$. Removing dependent paths ensures that the matrix $\textbf{\text{A}}^{T_iT_j}$ becomes full-rank, which guarantees a non-zero determinant for $\textbf{\text{A}}^{T_iT_j}$ and, consequently, for $\boldsymbol{\Sigma}^{T_iT_j}$.

The path-edge incidence matrix for the simple network structure in the second case of our first toy example in Figure \ref{fig:fig_3}(b) is
\[
    \text{\textbf{C}}^{T_2T_3} = \begin{array}{r@{}c@{}l}
    & \begin{array}{ccc}
        \pi_1^{T_2T_3} & \pi_2^{T_2T_3} & \pi_3^{T_2T_3}
      \end{array} \\
    \begin{array}{r}
        _{T_2T_3} \\ 
        _{T_2T_1} \\ 
        _{T_1T_3} \\
        _{T_1T_4} \\
        _{T_4T_3} \\
    \end{array} &
    \left( \begin{array}{c@{\hspace{22pt}}c@{\hspace{20pt}}c}
    1 & 0 & 0 \\
    0 & 1 & 1 \\
    0 & 1 & 0 \\
    0 & 0 & 1 \\
    0 & 0 & 1 \\
    \end{array} \right)_{5 \times 3} &
    \end{array}.
\]
For this example, we considered the same variance for all the comparisons ($\text{var}^{T_iT_j} = \sigma^2 = 0.3^2$); Therefore, the edge variance matrix is
  \[
  \text{\textbf{V}} =
 \begin{pmatrix}
     \text{var}^{T_2T_3} & 0 & 0 & 0 & 0 \\
    0 & \text{var}^{T_2T_1} & 0 & 0 & 0 \\
    0 & 0 & \text{var}^{T_1T_3} & 0 & 0 \\
    0 & 0 & 0 & \text{var}^{T_1T_4} & 0  \\
    0 & 0 & 0 & 0 & \text{var}^{T_4T_3}
 \end{pmatrix}_{5 \times 5} = 
  \begin{pmatrix}
    \sigma^2 & 0 & 0 & 0 & 0 \\
    0 & \sigma^2 & 0 & 0 & 0 \\
    0 & 0 & \sigma^2 & 0 & 0 \\
    0 & 0 & 0 & \sigma^2 & 0  \\
    0 & 0 & 0 & 0 & \sigma^2
 \end{pmatrix}_{5 \times 5}.
 \]
 Entering this $\text{\textbf{V}}$ matrix and  $\text{\textbf{C}}^{T_2T_3}$ into Equation \ref{eq:Sig}, we obtain the invertible variance-covariance matrix shown in Equation \ref{eq:covBC}.

\section{Reducing a Rank-Deficient Matrix to Its Rank \label{app1.2a}}

A square matrix \footnote{The concept of 'rank' is generally defined for any matrix with dimensions $D \times D^{\prime}$, but here, we only focus on square matrices as the path-adjacency and variance-covariance matrices are always symmetrical square matrices.} is called to be rank-deficient if its rows (or columns) are linearly dependent. In another words, the dimension of the matrix is greater than its rank. In this section, we use our second toy example in Figure \ref{fig:fig_4} to explain how one can reduce a rank-deficient matrix and obtain a full rank matrix. 

For the two treatments $T_1$ and $T_3$ in the toy example in Figure \ref{fig:fig_4}, five paths are detected via the depth-first search.\cite{DFS:book} These paths are shown in Figure \ref{fig:fig_5}. We create the path-adjacency matrix, $\text{\textbf{A}}^{T_1T_3}$ as 
\begin{equation}
    \text{\textbf{A}}^{T_1T_3} = \begin{array}{r@{}c@{}l}
    & \begin{array}{ccccc}
        \pi_1^{T_1T_3} & \pi_2^{T_1T_3} & \pi_3^{T_1T_3} & \pi_4^{T_1T_3} & \pi_5^{T_1T_3}
      \end{array} \\
    \begin{array}{r}
        \pi_1^{T_1T_3} \\ 
        \pi_2^{T_1T_3} \\ 
        \pi_3^{T_1T_3} \\
        \pi_4^{T_1T_3} \\
        \pi_5^{T_1T_3} \\
    \end{array} &
    \left( \begin{array}{c@{\hspace{22pt}}c@{\hspace{20pt}}c@{\hspace{20pt}}c@{\hspace{20pt}}c}
        2 & 1 & 1 & 0 & 0 \\
        1 & 3 & 0 & 2 & 1 \\
        1 & 0 & 3 & 2 & 1 \\
        0 & 2 & 2 & 4 & 2 \\
        0 & 1 & 1 & 2 & 3 \\
    \end{array} \right). &
    \end{array}
    \label{eq:AC}
\end{equation}
The diagonal elements in $\text{\textbf{A}}^{T_iT_j}$ represent the number of edges in that path. The off-diagonal elements represent the number of edges that the two paths share.  $\text{\textbf{A}}^{T_iT_j}$ matrix is the product of the $\text{\textbf{Z}}.\text{\textbf{Z}}^{\top}$, where $\text{\textbf{Z}}$ is the path-design matrix in R{\"u}cker et al.\cite{Gerta:shortest-path}

To reduce a rank-deficient matrix to its rank, we employ the Row Echelon Form (REF) method,\cite{golbert_REF} which is favored for its efficiency and numerical stability, particularly useful in handling large matrices and computational tasks. The method identifies linearly dependent rows which are removed along with their equivalent columns to create a square full rank matrix. The algorithms that perform the Echelon Form method can be found in the literature, such  as those described by Lay.\cite{echelonForm} Variations of this method are also available in many functions in {\tt R}   and {\tt Python}   including the {\tt rref()}   function from {\tt Python}   package {\tt sympy},\cite{sympy_package} or {\tt fullrank()}   function from {\tt R}   package {\tt robustbase},\cite{robustbase_package} or even {\tt qr()}   function from the base {\tt R}  . We use all of these functions to ensure that the following step-by-step illustration is not dependent on any specific programming function.

Denoting every row in Matrix \ref{eq:AC} as $r_i$, the following operations are done step by step to transform $ \text{\textbf{A}}^{T_1T_3}$ to $\Tilde{\text{\textbf{A}}}^{T_1T_3}$ in \ref{eq:M-transformed}:
\begin{align*}
\text{steps} \hspace{2cm} & \text{operation} \\
  s_1 \hspace{2cm} r_2 &= -\frac{1}{2}r_1 + r_2 \\
  s_2 \hspace{2cm} r_3 &= -\frac{1}{2}r_1 + r_3 \\
  s_3 \hspace{2cm} r_3 &= \frac{1}{5}r_2 + r_3 \\
  s_4 \hspace{2cm} r_4 &= -\frac{4}{5}r_2 + r_4 \\
  s_5 \hspace{2cm} r_5 &= -\frac{2}{5} r_2 + r_5 \\
  s_6 \hspace{2cm} r_4 &= -r_3 + r_4 \\
  s_7 \hspace{2cm} r_5 &= -\frac{1}{2}r_3 + r_5, 
\end{align*}
\begin{equation}
    \Tilde{\text{\textbf{A}}}^{T_1T_3} = \begin{pmatrix}
        2 & 1 & 1 & 0 & 0 \\
        0 & \frac{5}{2} & -\frac{1}{2} & 2 & 1 \\
        0 & 0 & \frac{12}{5} & \frac{12}{5} & \frac{6}{5} \\
        \textcolor{red}{0} & \textcolor{red}{0} & \textcolor{red}{0} & \textcolor{red}{0} & \textcolor{red}{0} \\
        0 & 0 & 0 & 0 & 2 \\
    \end{pmatrix}_{5 \times 5}.
    \label{eq:M-transformed}
\end{equation}
To create the REF, one must have the row(s) of zeroes at the bottom of the matrix (to get the so-called echelon shape). Therefore, the echelon form of $\text{\textbf{A}}^{T_1T_3}$  will be one step further (step $s_8$) by swapping the fourth with the fifth row. This yields 
\begin{equation}
        \text{REF}(\text{\textbf{A}}^{T_1T_3}) = \begin{pmatrix}
        2 & 1 & 1 & 0 & 0 \\
        0 & \frac{5}{2} & -\frac{1}{2} & 2 & 1 \\
        0 & 0 & \frac{12}{5} & \frac{12}{5} & \frac{6}{5} \\
        0 & 0 & 0 & 0 & 2 \\
        \textcolor{red}{0} & \textcolor{red}{0} & \textcolor{red}{0} & \textcolor{red}{0} & \textcolor{red}{0} \\
    \end{pmatrix}_{5 \times 5}.
    \label{eq:rowEchelonForm}
\end{equation}

A full row of zeroes in the row echelon form in \ref{eq:rowEchelonForm} represents  a deficiency in rank by one dimension for our path-adjacency matrix, $\text{\textbf{A}}^{T_1T_3}$. We keep the track of the indices of the rows that are swapped. Looking at $\Tilde{\text{\textbf{A}}}^{T_1T_3}$ in \ref{eq:M-transformed}, we observe that the fourth row is a full row of $0$s, which means the fourth path does not provide distinctive information. This indicates that the fourth row (path) can be expressed as a linear combination of other rows (paths). Therefore, removing this row and the corresponding fourth column does not diminish any information provided by the matrix. By removing the fourth row and column, now, we have a $4 \times 4$  full-rank matrix as
\begin{equation}
    \text{\textbf{A}}^{T_1T_3} = \begin{array}{r@{}c@{}l}
    & \begin{array}{ccccc}
        \pi_1 & \pi_2 & \pi_3 &  \pi_4 & \pi_5
      \end{array} \\
    \begin{array}{r}
        \pi_1 \\ 
        \pi_2 \\ 
        \pi_3 \\
        \pi_4 \\
        \pi_5 \\
    \end{array} &
    \left( \begin{array}{ccccc}
        2 & 1 & 1 & \textcolor{red}{0} & 0 \\
        1 & 3 & 0 & \textcolor{red}{2} & 1 \\
        1 & 0 & 3 & \textcolor{red}{2} & 1 \\
        \textcolor{red}{0} & \textcolor{red}{2} & \textcolor{red}{2} & \textcolor{red}{4} & \textcolor{red}{2} \\
        0 & 1 & 1 & \textcolor{red}{2} & 3 \\
    \end{array} \right) &
    \end{array} \quad \rightarrow \quad
 \text{\textbf{A}}^{T_1T_3} = \begin{pmatrix}
        2 & 1 & 1 &  0 \\
        1 & 3 & 0 &  1 \\
        1 & 0 & 3 &  1 \\
        0 & 1 & 1 &  3 \\
\end{pmatrix}_{4 \times 4}
\label{eq:independentMatrix}.
\end{equation}  

According to the reduced $\text{\textbf{A}}^{T_1T_3}$ in (\ref{eq:independentMatrix}), we now have $\bm{P}^{\prime} = 4$ linearly independent paths. Therefore, we write  the variance-covariance matrix for this example as
\begin{equation}
     \boldsymbol{\Sigma}^{T_1T_3} = \begin{pmatrix}
        \text{var}^{T_1T_2} + \text{var}^{T_2T_3} &  \text{var}^{T_1T_2} & \text{var}^{T_2T_3} &  0 \\
        \text{var}^{T_1T_2} & \text{var}^{T_1T_2} + \text{var}^{T_2T_4} + \text{var}^{T_4T_3} & 0 &  \text{var}^{T_4T_3} \\
        \text{var}^{T_2T_3} & 0 & \text{var}^{T_1T_5} + \text{var}^{T_5T_2} + \text{var}^{T_2T_3} &  \text{var}^{T_1T_5} \\
        0 & \text{var}^{T_4T_3} & \text{var}^{T_1T_5} &  \text{var}^{T_1T_5} + \text{var}^{T_5T_4} + \text{var}^{T_4T_3} \\
\end{pmatrix}_{4 \times 4}.
\end{equation}

\section{Different Sets of Linearly Independent Paths \label{app1.3a}}

The REF of a matrix, if it exists, is not unique. A rank-deficient matrix can have multiple echelon forms depending on its linearly dependent rows. In $\text{\textbf{A}}^{T_1T_3}$,  $r_1$,  $r_2$, $r_3$, and $r_4$ are linearly dependent. Therefore, there are sequences of steps and operations that  result in an $\Tilde{\text{\textbf{A}}}^{T_1T_3}$ with a full row of zeroes appearing in the first row, the second row, or the third row. This implies that  any of the rows $r_1$,  $r_2$, $r_3$, or $r_4$, along with their corresponding columns, can be removed to obtain a full-rank matrix. This reflects the concept described in Equation \ref{eq:linCombo}, which allowed us to remove any of the paths $\pi_1^{T_1T_3}$, $\pi_2^{T_1T_3}$, $\pi_3^{T_1T_3}$, or $\pi_4^{T_1T_3}$ to form a set of linearly independent paths. This indicates that different full-rank variance-covariance matrices can be generated depending on the specific row(s) removed.   We use Gaussian elimination to perform row operations that transform the initial matrix into one of the following matrices:
\begin{equation}
\begin{aligned} 
    \Tilde{\text{\textbf{A}}}^{T_1T_3}_1 &= \begin{pmatrix}
        &\textcolor{red}{0} & \textcolor{red}{0} & \textcolor{red}{0} & \textcolor{red}{0} & \textcolor{red}{0} \\
        &0   &2.5 &-0.5  &2   &1 \\
        &0   &2.5 &-0.5  &2   &1 \\
        &0   &2 &2  &4   &2 \\
        &0   &1 &1  &2   &3
    \end{pmatrix},  \quad   \Tilde{\text{\textbf{A}}}^{T_1T_3}_2 &= \begin{pmatrix}
        &\frac{5}{3}   &0   &1   &\frac{-2}{3}   &\frac{-1}{3} \\
        &\textcolor{red}{0} & \textcolor{red}{0} & \textcolor{red}{0} & \textcolor{red}{0} & \textcolor{red}{0} \\
        &1   &0 &3  &2   &1 \\
        &\frac{-2}{3}   &0 &2  &\frac{8}{3}   &\frac{4}{3} \\
        &\frac{-1}{3}   &0 &1  &\frac{4}{3}   &\frac{8}{3}
    \end{pmatrix}, \\
    \Tilde{\text{\textbf{A}}}^{T_1T_3}_3 &= \begin{pmatrix}
        &\frac{5}{3}   &1   &0   &\frac{-2}{3}   &\frac{-1}{3} \\
        &1   &3   &0   &2   &1 \\
        &\textcolor{red}{0} & \textcolor{red}{0} & \textcolor{red}{0} & \textcolor{red}{0} & \textcolor{red}{0} \\
        &\frac{-2}{3}   &2 &0  &\frac{8}{3}   &\frac{4}{3} \\
        &\frac{-1}{3}   &1 &0  &\frac{4}{3}   &\frac{8}{3}
    \end{pmatrix}, \quad     \Tilde{\text{\textbf{A}}}^{T_1T_3}_4 &= \begin{pmatrix}
        &2   &1   &1   &0   &0 \\
        &1   &2   &-1   &0   &0 \\
        &1   &-1   &2   &0   &0 \\
        &\textcolor{red}{0} & \textcolor{red}{0} & \textcolor{red}{0} & \textcolor{red}{0} & \textcolor{red}{0} \\
        &0   &0   &0   &0   &2
    \end{pmatrix}.
    \end{aligned}
    \label{eq:fourRowZeroes}
\end{equation}
Each of these matrices corresponds to a specific REF.

\begin{figure}[h!]
    \centering
        \includegraphics[width=0.8\textwidth]{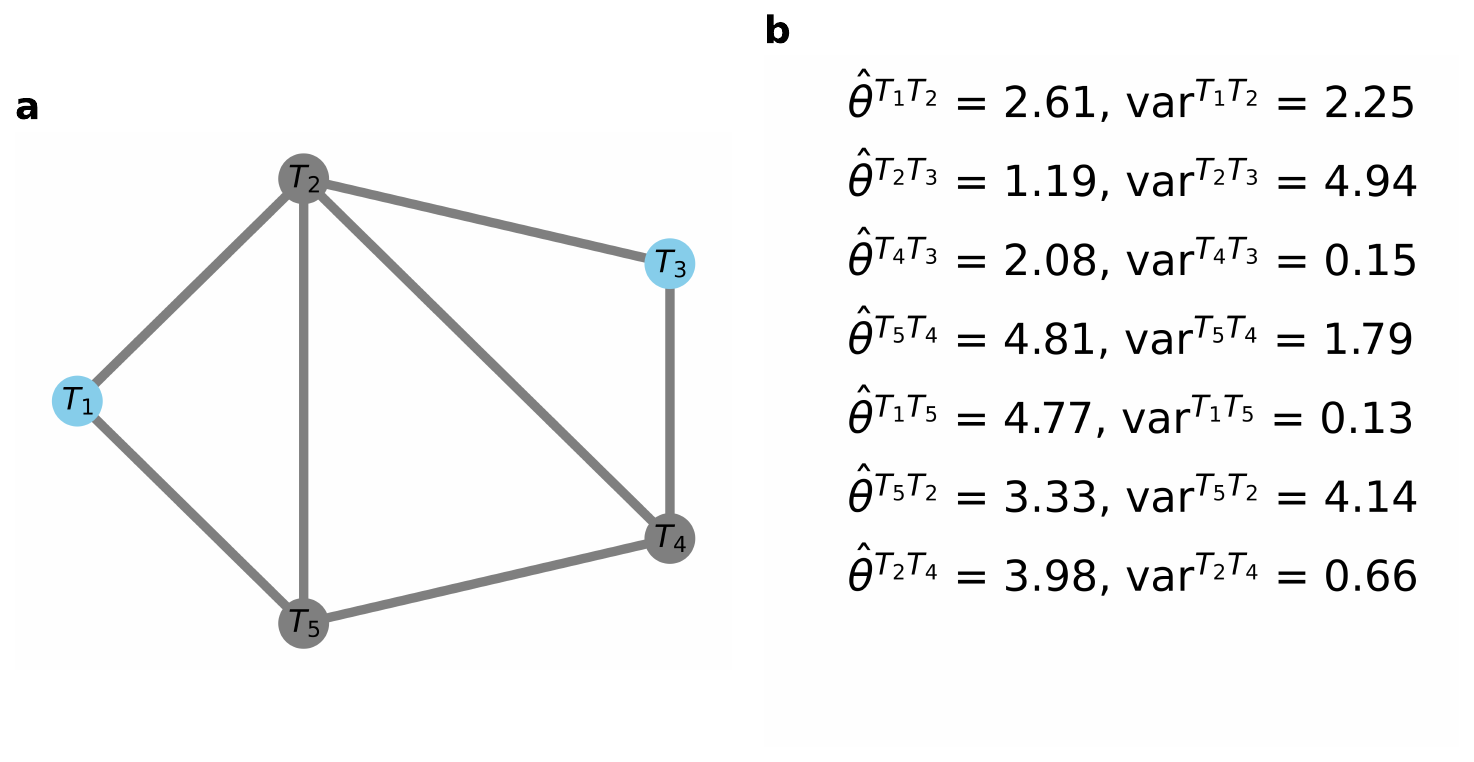}  
        \caption{Fictional example in Figure \ref{fig:fig_4} with arbitrary values for the effects and variances for the seven direct comparisons.}
        \label{fig:app}
\end{figure}

An important  observation is that no matter which row is removed, the value of \( Q^{\text{path}}_{T_iT_j} \) remains unchanged. This is because as the row and the corresponding column are removed from the variance-covariance matrix, the respective row is also removed from the path and NMA estimate vectors. This means that the choice of linearly independent paths does not affect the results or inferences related to the measure of inconsistency.  An illustrative explanation for this is given below for the toy example in Figure \ref{fig:fig_4}. We consider the  arbitrary  values for the effects and variances of the direct comparisons shown in Figure \ref{fig:app}. 

Here, the comparison of interest is $T_1T_3$, where four out of five paths are linearly dependent according to  Equation \ref{eq:linCombo}. Extensive but easy calculations  show that the \( Q^{path}_{T_1T_3}\) is the same if any of these four paths is removed from the calculations.  This means that it does not matter which path is removed, the chosen set of linearly independent paths will give us the same result, with $ Q^{path}_{T_1T_3} = 3.32$. Since the DOF is $4$ in all cases, it will result in the same  p-value for the inconsistency of $T_1$ vs $T_3$ ($\text{p-value} = 0.5$). 

\begin{figure}[h!]
    \centering
    \includegraphics[width=0.5\linewidth]{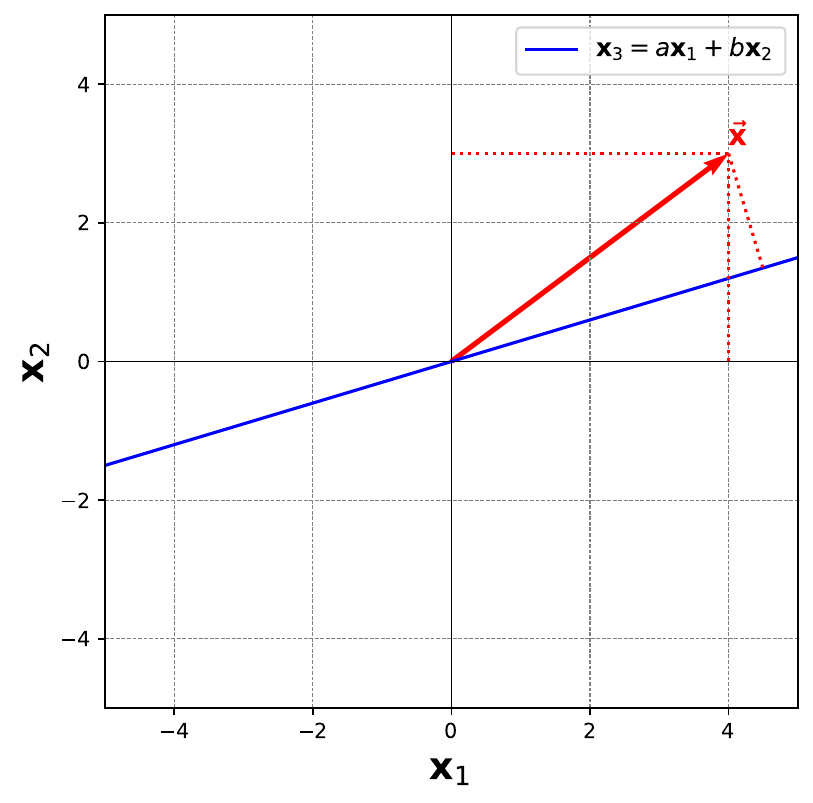}
    \caption{Representation of a vector, $\vec{\textbf{x}},$ defined by three vectors\textendash $\textbf{x}_1$, $\textbf{x}_2$, and $\textbf{x}_3$—in a $2$D Cartesian space. The red dotted lines, show the projections of $\vec{\textbf{x}}$ onto these three axes. This  does not alter the magnitude of the vector in space.}
    \label{fig:fig_10}
\end{figure}

It is noteworthy that when the rows of a square matrix are linear combinations of other rows, it indicates that the matrix has more rows than the dimension of its row space, which means we are defining the space with more rows than needed. This does not affect the norm or magnitude of a vector in this space. If we have a transformation matrix \( \text{\textbf{M}} \), the norm or magnitude of a vector \( \vec{\textbf{x}} \) is defined as \( \vec{\textbf{x}}^{\top} \text{\textbf{M}} \vec{\textbf{x}} \).  This implies that \( Q^{\text{path}}_{T_iT_j} \)\textendash representing the norm or magnitude of the effect deviations, \( (\boldsymbol{\Theta}(\pi^{T_iT_j}) - \hat{\boldsymbol{\Theta}}^{T_iT_j(\text{nma})}) \)—remains unchanged whether all paths or only the linearly independent paths (the basis vectors in our path space) are used. This concept is illustrated in Figure \ref{fig:fig_10}. In a $2$D Cartesian space, \( \textbf{x}_1 \) and \( \textbf{x}_2 \) are  two linearly independent basis vectors. Now, if the transformation matrix is extended to include a third dimension, represented by the equation \( \textbf{x}_3 = a\textbf{x}_1 + b\textbf{x}_2  \), the norm or magnitude of the vector \( \vec{\textbf{x}} \) remains unaffected.
This implies that we can use the rank-deficient \( \text{\textbf{A}}^{T_iT_j}_{\bm{P} \times\bm{P}} \) in \ref{eq:AC} and still obtain the same value for our measure of inconsistency, $Q^{\text{path}}_{T_iT_j}$. For the example discussed in this section, we find \( Q^{\text{path}}_{T_1T_3} = 3.32 \), even without removing any row or column from $\boldsymbol {\Sigma}^{T_1T_3}$. This demonstrates that the same value for \( Q^{\text{path}}_{T_iT_j} \) can be obtained using all paths of evidence between two treatments, \( T_i \) and \( T_j \).

However, there are two important considerations to keep in mind: (1) Understanding the number of linearly independent paths remains essential for determining the DOF. This DOF is crucial for performing a statistical test and obtaining a p-value for the null hypothesis; And (2) When using all paths, the variance-covariance matrix is no longer invertible (\( \text{det}(\boldsymbol{\Sigma}^{T_iT_j}) = 0 \)). Consequently, a pseudo-inverse method must be employed to approximate \( (\boldsymbol{\Sigma}^{T_iT_j})^{-1} \).\cite{psudo_inverse}


\newpage
\setcounter{section}{0}
\renewcommand{\thesection}{S\arabic{section}} 
\setcounter{MaxMatrixCols}{12}
\section*{SUPPORTING INFORMATION} 

\section{Software}
Our path-based method has been implemented in the \texttt{netmeta} package in \texttt{R}. It is accessible through the \texttt{netpath()} function. If the dataset is for the aggregate network, the mandatory arguments are the first and second treatments of interest. For the first case in our first toy example in Figure 2(b) in the main text, the function call will be: \texttt{netpath(node1 = "T\_1", node2 = "T\_3")}, where the output is:
\begin{verbatim} 
Comparison   Q    p_value     No. of independent paths 
T_1:T_3     11.11  0.003          3
\end{verbatim}
The output consists of four columns: the first column lists the chosen comparison, while the other three columns report the corresponding $Q^{\text{path}}_{T_iT_j}$, p-value, and the number of independent paths for each comparison. The degree of freedom for each comparison is one less than the number of independent paths. 
If the dataset is in study–level format (not aggregate), one must also provide a \texttt{netmeta} object, \texttt{nma\_obj}. The function call will then be:
\texttt{netpath(nma\_obj, node1 = "T\_1", node2 = "T\_3")}, with the same output. 
 
 For a specific comparison of interest, \texttt{netpath\_plot} is an optional argument that saves a Netpath plot like Figures (7) and (9) in the main text in the defined directory. The \texttt{verbose} is another optional argument with the default to be set to \texttt{FALSE}.  When enabled, it provides additional information such as:
\begin{enumerate}
    \item hat matrix, $\textbf{\text{H}}$
    \item path-adjacency matrix, $\textbf{\text{A}}$
    \item variance–covariance matrix, $\boldsymbol{\Sigma}$
    \item a complete set of all paths detected between the two treatments. For each path, it prints the sequence of nodes from \texttt{node1} to \texttt{node2}, the size of the path (i.e., the number of edges), and the total effect and variance of that path
    \item the paths removed from the $Q$ calculation due to linear dependency
\end{enumerate}
For $T_1T_3$ comparison in the arbitrary network in Section A.3 of the Appendix, it will show:
        \begin{verbatim}
The total number of paths detected between treatment  1  and treatment  3  is  5 
path # 1  : {T_1, T_2, T_3}
 size: 2     total effect: 3.8   total variance: 3.69
path # 2  : {T_1, T_2, T_4, T_3}
 size: 3     total effect: 8.67   total variance: 3.38  
path # 3  : {T_1, T_5, T_2, T_3}
 size: 3     total effect: 9.29   total variance: 5.53  
path # 4  : {T_1, T_5, T_2, T_4, T_3}
 size: 4     total effect: 14.16  total variance: 5.22  
path # 5  : {T_1, T_5, T_4, T_3}
 size: 3     total effect: 11.66  total variance: 2.27  
The following paths are removed from calculation due to linear dependency: 
  path #4
\end{verbatim}
If one wishes to retrieve only the independent paths and their respective statistics, the optional argument \texttt{indep\_stat} must be set to \texttt{TRUE}, as its default value is \texttt{FALSE}.

\section{Paths in the Real-World Example} \label{sec:paths_realEx}
For our real-world example in Section 4 of the main text, we use the data for the $11$ treatments for depression provided by Linde et al. \cite{depressionData}. First, we use the \texttt{netmeta} package in \texttt{R} to call \texttt{netmeta()}. This creates a \texttt{netmeta} object, \texttt{nma\_obj}, which we use to construct the $\textbf{\text{H}}$ matrix by calling \texttt{hatmatrix(nma\_obj, method="Davies", type="full")}, giving us a $55 \times 55$ full $\textbf{\text{H}}$ matrix. We retrieve the $T_2T_5$ row using the command \texttt{H[["common"]]["2:5",]}.  Using this row and following Section 2 of the main text, we create the directed network shown in Figure 8(b) of the main text. We then use \texttt{netpath(nma\_obj, node1 = "T\_2", node2 = "T\_5")} to perform a depth-first search algorithm and obtain a full list of $49$ paths, along with the respective statistics on their size, total effect, and variance. However, there are only $11$ linearly independent paths from $T_2$ to $T_5$. These paths are detected by \texttt{netpath()} and can be printed as output by setting \texttt{indep\_stat=TRUE}. These paths and their respective statistics are as follows:
\begin{verbatim}
path # 1  : {2 6 1 3 5          }
 size: 4     total effect: 1.370665   total variance: 0.382586  
path # 2  : {2 6 1 9 3 5        }
 size: 5     total effect: 1.042525   total variance: 0.371354  
path # 3  : {2 6 1 9 4 3 5      }
 size: 6     total effect: 1.136702   total variance: 0.499311  
path # 4  : {2 6 1 9 5          }
 size: 4     total effect: 0.877207   total variance: 0.275334  
path # 5  : {2 6 3 5            }
 size: 3     total effect: 1.274425   total variance: 0.297619  
path # 6  : {2 6 7 1 3 5        }
 size: 5     total effect: 1.500269   total variance: 0.819754  
path # 7 : {2 6 7 9 3 5        }
 size: 5     total effect: 0.446684   total variance: 0.747420 
path # 8 : {2 6 9 3 5          }
 size: 4     total effect: 0.562831   total variance: 0.197955 
path # 9 : {2 8 6 1 3 5        }
 size: 5     total effect: 1.398439   total variance: 0.421127  
path # 10 : {2 11 1 3 5         }
 size: 4     total effect: -0.068064  total variance: 0.578129 
path # 11 : {2 11 6 1 3 5       }
 size: 5     total effect: 0.954114   total variance: 0.479962  
\end{verbatim}
According to these $11$ independent paths, the path effect estimate vector for this example will be:
\[
\left(\boldsymbol{\hat{\Theta}}^{T_2T_5}\right)^\top = 
\begin{bmatrix}
    1.37 & 1.04 & 1.13 & 0.87 & 
    1.27 & 1.50 & 0.44 & 0.56 & 
    1.39 & -0.06 & 0.95
\end{bmatrix}_{1 \times 11}.
\]
The corresponding covariance matrix is
\[
\boldsymbol{\Sigma}^{T_2T_5} =
\begin{pmatrix}
0.38 & 0.29 & 0.29 & 0.16 & 0.16 & 0.25 & 0.16 & 0.16 & 0.36 & 0.23 & 0.36 \\
0.29 & 0.37 & 0.36 & 0.22 & 0.16 & 0.16 & 0.17 & 0.17 & 0.28 & 0.14 & 0.28 \\
0.29 & 0.36 & 0.50 & 0.22 & 0.16 & 0.16 & 0.16 & 0.16 & 0.28 & 0.14 & 0.28 \\
0.16 & 0.22 & 0.22 & 0.28 & 0.02 & 0.02 & 0.02 & 0.02 & 0.14 & 0.00 & 0.14 \\
0.16 & 0.16 & 0.16 & 0.02 & 0.30 & 0.16 & 0.16 & 0.16 & 0.14 & 0.14 & 0.14 \\
0.25 & 0.16 & 0.16 & 0.02 & 0.16 & 0.82 & 0.45 & 0.16 & 0.23 & 0.23 & 0.23 \\
0.16 & 0.17 & 0.16 & 0.02 & 0.16 & 0.45 & 0.75 & 0.17 & 0.14 & 0.14 & 0.14 \\
0.16 & 0.17 & 0.16 & 0.02 & 0.16 & 0.16 & 0.17 & 0.20 & 0.14 & 0.14 & 0.14 \\
0.36 & 0.28 & 0.28 & 0.14 & 0.14 & 0.23 & 0.14 & 0.14 & 0.42 & 0.23 & 0.36 \\
0.23 & 0.14 & 0.14 & 0.00 & 0.14 & 0.23 & 0.14 & 0.14 & 0.23 & 0.58 & 0.29 \\
0.36 & 0.28 & 0.28 & 0.14 & 0.14 & 0.23 & 0.14 & 0.14 & 0.36 & 0.29 & 0.48 \\
\end{pmatrix}_{11 \times 11}.
\]

These outputs are accessible via \texttt{netpath()} when \texttt{verbose} is \texttt{TRUE}.

\end{document}